\documentclass{jfm}
\usepackage{graphicx}
\usepackage{epstopdf, epsfig}

\usepackage{subcaption}
\usepackage{enumitem}

\usepackage{url}
\usepackage{amsmath}
\usepackage{amssymb}
\renewcommand{\vec}[1]{\mathbf{#1}} 
\newcommand{\bs}[1]{\boldsymbol{#1}} 
\renewcommand{\Re}{\Rey} 
\newcommand{\Rep}{\Rey_{p}} 

\newcommand{\cross}{{\times}} 

\graphicspath{{./figures/}} 

\shorttitle{Inertial focusing in curved ducts at moderate flow rate}
\shortauthor{B. Harding and Y. M. Stokes}
\title{Inertial focusing of spherical particles in curved microfluidic ducts at moderate Dean numbers}

\author{
  Brendan Harding\aff{1}
  \corresp{\email{brendan.harding@adelaide.edu.au}},
  \and Yvonne M. Stokes\aff{2}
}

\affiliation{
  \aff{1}School of Mathematics and Statistics, Victoria University Wellington,
  Wellington 6140, New Zealand
  \aff{2}School of Mathematical Sciences, The University of Adelaide,
  Adelaide, South Australia 5005, Australia
}

\begin{document}

\maketitle

\abstract{
We examine the effect of Dean number on the inertial focusing of spherical particles suspended in flow through curved microfluidic ducts.
Previous modelling of particle migration in curved ducts assumed the flow rate was small enough that a leading order approximation of the background flow with respect to the Dean number produces a reasonable model.
Herein, we extend our model to situations having a moderate Dean number (in the microfluidics context) while the particle Reynolds number remains small.
This extension allows us to capture changes in the background flow that occur with increasing flow rate, namely a shift in local extrema towards the outside wall.
The change in the axial velocity profile of the background flow has an effect on the inertial lift force, while the change in the cross-sectional components directly affects the secondary flow drag.
In keeping the particle Reynolds number small we approximate the inertial lift force in a similar manner to previous studies while capturing subtle effects do to the modified background flow profile.
Capturing and understanding these effects is an important step towards accurately modelling inertial migration across a wide range of practical applications.
Our results reveal how the changing background flow profile modifies the inertial focusing of particles. 
We illustrate enhanced lateral separation of particles by size in a number of scenarios and find that focusing times can be roughly separated into two regimes. 
These results suggest our model might aid with parameter choices for separation of particles by size.
}

\section{Introduction}\label{sec:intro}

Inertial focusing of particles suspended in flow through curved and spiral microfluidic ducts has been studied extensively in the experimental literature, particularly in relation to its application to size based particle/cell separation~\citep{SeoEtal2007,BhagatEtal2008,DiCarlo2009,MartelToner2012,Wetal2014,WarkianiEtal2016,RafeieEtal2019}.
Much is known about the nature of the inertial lift force in a variety of situations involving a particle suspended in flow between two plane parallel walls~\citep{Saffman1965,HoLeal1974,SchonbergHinch1989,Hogg1994,Asmolov1999}. 
However, the nature of the inertial lift force is very different for a fully enclosed duct making these studies of limited use in understanding practical applications.
Sufficiently small particles suspended in flow through straight ducts with square cross-section are known to focus at one of four equilibria located a small distance from the centre of each side wall~\citep{DiCarloEtal2009,HoodLeeRoper2015}.
This is also the case for straight rectangular ducts, although stable equilibria near the shorter side walls attract relatively few particles and disappear entirely for larger particles~\citep{MartelToner2013,HoodThesis2016}.
In curved ducts the migration of particles becomes complicated due to the secondary vortices that are generated as part of the Dean flow through the duct.
The interaction of the drag force from these vortices with the inertial lift force leads to a wide variety of particle migration dynamics~\citep{GossettDiCarlo2009,MartelToner2014,HaEtal22,ValaniEtal22}.

Our previous study conducted a detailed examination of the migration of neutrally buoyant spherical particles suspended in a sufficiently slow flow through curved rectangular ducts \citep{HardingStokesBertozzi2019}.
An accurate model of particle migration was developed by coupling the particle motion to a Navier--Stokes model of the fluid flow.
By using a carefully chosen reference frame, and applying a suitable non-dimensionalisation and perturbation expansion of both the background and disturbance flows, the individual forces primarily responsible for driving particle migration were separated and then estimated via numerical simulation.
These forces were then re-assembled into a system of ordinary differential equations to model particle trajectories.
It was found particles migrated towards stable equilibria whose horizontal location approximately collapsed onto a single curve when plotted against the parameter $\kappa=\ell^4/(4a^3R)$, with $\ell$ being the duct height, $a$ being the particle radius, and $R$ being the bend radius.
It was later shown how non-neutrally buoyant particles could be modelled by adding suitable perturbations to the neutrally buoyant model \citep{HardingStokes2020}.

A key part of the prior modelling was an assumption that the flow rate is small enough that both $\Rep=\Rey(a/\ell)^2$ and $K=\epsilon\Rey^2$ are small, where $\epsilon=\ell/(2R)$ and $\Rey=(\rho/\mu)U_m(\ell/2)$ is the channel Reynolds number with $\rho$ being the fluid density, $\mu$ the fluid viscosity and $U_m$ the maximum velocity of the background flow down the main axis.
This allowed us to take the leading order contribution of each perturbation expansion as a reasonable approximation. 
In a typical practical setting, these assumptions only hold when $\Rey\lesssim\mathcal{O}(10)$.
In contrast, most microfluidics experiments in the literature correspond to $\Rey=\mathcal{O}(100)$.
While we expect our existing model may still have qualitative value at these flow rates, they are of less use quantitatively.

Substantially higher flow rates, e.g. $\Rey\gtrsim\mathcal{O}(1000)$, are of limited practical interest for a couple of reasons.
The first is that the increasing strength of the Dean flow eventually inhibits the ability of particles to focus, as seen in the decreasing sharpness factor with increasing flow rate in the experimental results of \citet{RafeieEtal2019}. 
Second, there is a critical Dean number, depending on the specific cross-section, above which the secondary component of the background flow exhibits multiple vortex pairs~\citep{Winters1987} and this seems generally undesirable for most applications.

In this paper we incorporate additional terms from the perturbation expansion of the background flow into the model. 
This has the effect of increasing the values of $K$ for which our model has quantitative value.
Since we continue to use a leading order approximation of the disturbance flow, this model doesn't expand the applicability in cases where the magnitude of $\Rep$ is the limiting factor (e.g. when the particle is relatively large).
Nonetheless, we feel this provides valuable insights and is an important step towards producing an accurate model applicable to a wide range of physical set ups.
This work also illustrates how the symmetry associated with a curved duct leads to a decoupling of axial and secondary parts in the disturbance flow at leading order which each generate inertial lift in their counterpart at first order.

The paper is organised as follows.
Section~\ref{sec:background} describes the general setup of the problem and briefly summarises the modelling of forces driving particle migration as developed in \citet{HardingStokesBertozzi2019}.
We also introduce some notation to support the remainder of the text and remark on the applicability of the Lorentz reciprocal theorem to torque calculations.
Section~\ref{sec:bg_flow} describes our improved approximation of the background flow which utilises multiple terms from a perturbation expansion with respect to the Dean number $K$.
Section~\ref{sec:qv0_expansion} describes how the new background flow approximation is incorporated into the inertial lift calculation and ultimately leads to a system of first order differential equations which describe particle migration.
Section~\ref{sec:results} reports a range of findings obtained from the new model: 
firstly, we examine how the horizontal location of stable equilibria are perturbed by increasing $K$;
secondly, we examine how the parameters $\epsilon,K,\kappa,\alpha(=2a/\ell)$ influence how long it takes particles to focus;
thirdly, we examine how increasing $K$ influences previously observed trends in the horizontal location of stable equilibria with respect to $\epsilon^{-1}$ and $\kappa$. 
Section~\ref{sec:conclusions} summarises our findings and remarks on avenues for future exploration.

\section{Problem setup and theoretical background}\label{sec:background}

\begin{figure} 
\centering
\includegraphics{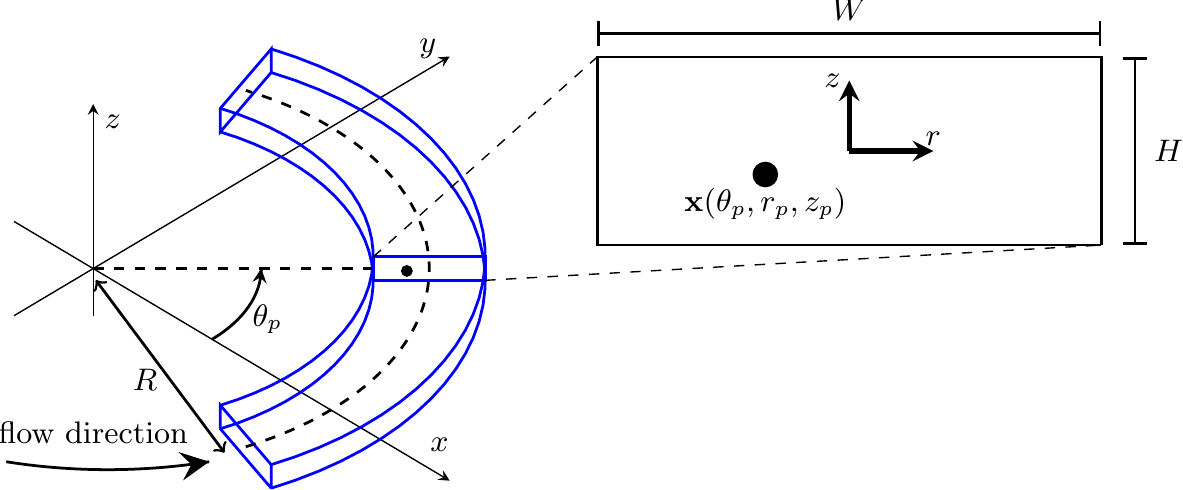}
\caption{
Curved duct with rectangular cross-section containing a spherical particle located at $\vec{x}_{p}=\vec{x}(\theta_{p},r_{p},z_{p})$. 
The enlarged view of the cross-section containing the particle illustrates the origin of the local $r,z$ coordinates at the centre of the duct.
The bend radius $R$ is with respect to the centre-line of the duct. 
Note that we do not consider the flow near the inlet/outlet.
Adapted from \citet{HardingStokesBertozzi2019}.}\label{fig:duct_setup}
\end{figure}

Our curved duct setup remains identical to that in \citet{HardingStokesBertozzi2019} and is depicted in Figure~\ref{fig:duct_setup}.
The (stationary) lab reference frame is $\mathbf{x}=x\mathbf{i}+y\mathbf{j}+z\mathbf{k}$ with the duct bending around the $z$-axis.
The duct domain is most readily described using the cylindrical coordinate system $(r,\theta,z)$ for which
\begin{equation}\label{eqn:cc}
\mathbf{x}(r,\theta,z) = (R+r)\cos(\theta)\mathbf{i}+(R+r)\sin(\theta)\mathbf{j}+z\mathbf{k} \,,
\end{equation}
where $R$ is the bend radius of the duct measured from the origin (of the lab frame) to the centre of the cross-section.
The cross-section itself is described by $(r,z)\in\mathcal{C}$ (with origin $(r,z)=(0,0)$ in the centre of the cross-section).
The duct interior is then described by $\mathcal{D}=\{\mathbf{x}(\theta,r,z) \mid (r,z)\in\mathcal{C}\}$.
While our approach can be applied to any desired cross-section $\mathcal{C}$, this study is concerned with rectangular cross-sections having width $W$ and height $H$, thus
\begin{equation}
\mathcal{C} = \left\{(r,z) :  |r|\leq W/2 \,,\, |z|\leq H/2 \right\}\,.
\end{equation}
We take $\ell=\min\{W,H\}$ to be the characteristic length scale of the duct.
Of principle interest will be ducts with $W\geq H$, and thus $\ell=H$, as these are most common in the experimental literature.

\begin{figure}
\centering
\includegraphics[scale=0.8]{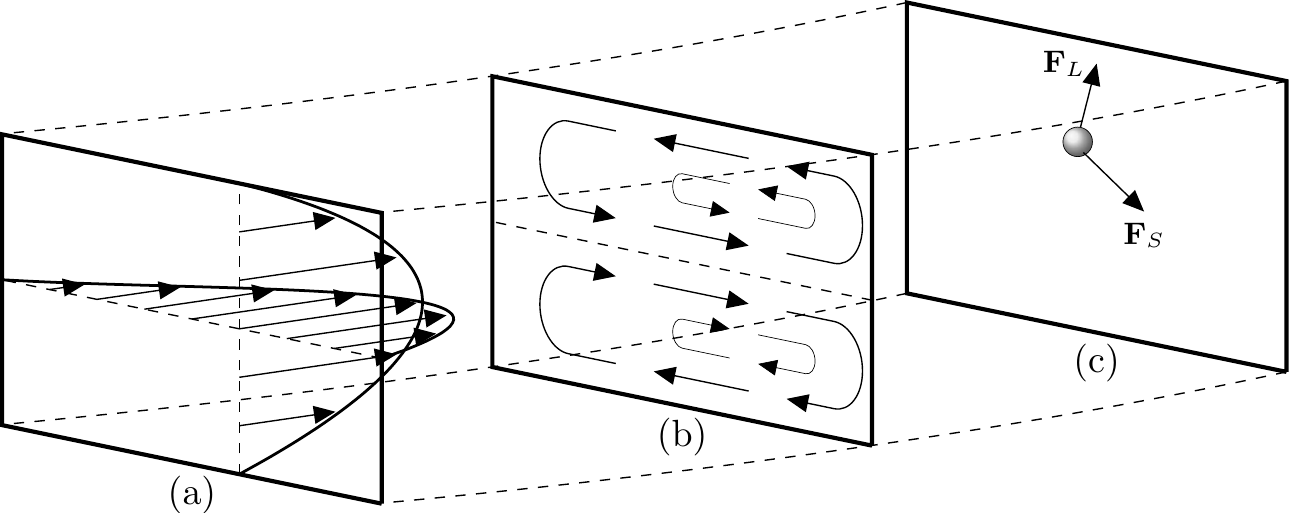}
\caption{
Cross-sections of a curved rectangular duct depicting: 
(a) The axial component of the background flow; 
(b) the secondary component of the background flow consisting of two vertically symmetric counter rotating vortices;
(c) a spherical particle and the primary cross-sectional forces which drive its migration. 
Here $\mathbf{F}_{S}$ is the drag from the secondary component of the background flow, and $\mathbf{F}_{L}$ is the inertial lift force. 
The magnitude and direction of each vector are for illustration only. 
Gravitational and centrifugal/centripetal forces are omitted.
The background flow is shown to be skewed towards the outside wall of the curved duct (here on the right), as is expected at moderate Dean numbers.
Adapted from \cite{HardingStokes2020}.}\label{fig:duct_setup2}
\end{figure}

Steady pressure driven flow through the duct (in the absence of any particles) is referred to as the background flow.
The fluid is assumed to be incompressible with constant density $\rho$ and viscosity $\mu$.
The pressure and velocity fields are denoted $\bar{p}$ and $\bar{\mathbf{u}}$, respectively, and are modelled using the Navier--Stokes equations.
We take $U_{m}$ to be a characteristic velocity of this flow, approximately describing the maximum axial velocity. 
The channel/duct Reynolds number is then $\Rey:=(\rho/\mu)U_{m}(\ell/2)$. 
Additionally, letting $\epsilon=\ell/(2R)$ denote the relative curvature, we define the Dean number as $K=\epsilon\Rey^{2}$ after \citet{Dean1927,DeanHurst1959} who studied the secondary flow which develops within curved duct flow.
The interaction of drag from this secondary flow with the inertial lift force, primarily due to the axial flow, produces a variety of particle migration dynamics that are exploited in a wide range of microfluidics applications.
The nature of the background flow and its approximation for the purposes of this study will be discussed further in Section~\ref{sec:bg_flow}.
Figure~\ref{fig:duct_setup2} depicts the axial (a) and secondary (b) components of the background flow, and (c) depicts the competing secondary drag and inertial lift forces on a particle.

Now, consider a single particle suspended in the flow.
Let $\mathcal{P}:=\{\vec{x}:|\vec{x}-\vec{x}_{p}|<a\}$ denote a spherical particle with constant density $\rho_{p}$, radius $a$ and centred at $\vec{x}_{p}(t)$ (such that $\mathcal{P}\subset\mathcal{D}$).
We let $r_p,\theta_p,z_p$ denote its cylindrical coordinates, i.e. $\mathbf{x}_{p}=\mathbf{x}(r_{p},\theta_{p},z_{p})$, where $r_{p},\theta_{p},z_{p}$ are each functions of $t$.
The particle travels with a velocity $\mathbf{u}_{p}(t):=d \mathbf{x}_{p}/d t$ and spins freely about its centre with angular velocity $\boldsymbol{\Omega}_{p}(t)$.
Thus, a point $\mathbf{x}$ within $\mathcal{P}$ has instantaneous velocity $\mathbf{u}_{p}+\boldsymbol{\Omega}_{p}\cross(\mathbf{x}-\mathbf{x}_{p})$.
The fluid domain is denoted $\mathcal{F}:=\mathcal{D}\backslash\mathcal{P}$ and its boundary $\partial\mathcal{F}$ consists of the duct walls $\partial\mathcal{D}$ and the particle surface $\partial\mathcal{P}$.
The fluid flow is now described by the pressure and velocity fields $p,\mathbf{u}$, respectively, which are also modelled by the Navier--Stokes equations, specifically
\begin{subequations}
\begin{align}
\nabla\cdot\sigma(p,\vec{u}) &= \rho\left(\frac{\partial\vec{u}}{\partial t}+\vec{u}\cdot\nabla\vec{u}-\vec{g}\right) &&\text{for $\vec{x}\in\mathcal{F}$,} \label{eqn:ns_a}\\
\nabla\cdot\vec{u} &= 0 &&\text{for $\vec{x}\in\mathcal{F}$,} \\
\vec{u} &= \vec{0} &&\text{for $\vec{x}\in\partial\mathcal{D}$,} \\ 
\vec{u} &= \vec{u}_{p}+\bs{\Omega}_{p}\cross(\vec{x}-\vec{x}_{p}) &&\text{for $\vec{x}\in\partial\mathcal{P}$} 
\end{align}
\end{subequations}
where $\sigma(p,\vec{u})$ is the stress tensor
\begin{equation}
\sigma(p,\vec{u}):=-p\mathbb{I}+\mu\left(\nabla\vec{u}+\nabla\vec{u}^{\intercal}\right) \,.
\end{equation}

The gravitational body force $\mathbf{g}$ is only important for non-neutrally buoyant particles ($\rho_{p}\neq\rho$).
In \citet{HardingStokes2020} it was demonstrated that with $\mathbf{g}=-g\mathbf{k}$ the influence of non-neutral buoyancy can be separated from the migration model and subsequently treated as an additional perturbation of the force experienced by a neutrally buoyant particle.
This remains true in this work so we simplify the development by considering a neutrally buoyant particle (i.e. $\rho_{p}=\rho$) and dropping the gravitational acceleration from \eqref{eqn:ns_a}.

The motion of the suspended neutrally buoyant particle is driven solely by the hydrodynamic force and torque exerted on it.
Specifically
\begin{subequations}\begin{align}
\mathbf{F} &:= \int_{\partial\mathcal{P}}(-\mathbf{n})\cdot\sigma(p,\mathbf{u})\,dS \,, \\
\mathbf{T} &:= \int_{\partial\mathcal{P}}(\mathbf{x}-\mathbf{x}_{p})\cross\left((-\mathbf{n})\cdot\sigma(p,\mathbf{u})\right)\,dS \,,
\end{align}\end{subequations}
describe the force and torque, respectively, where $\mathbf{n}$ is taken to be the normal with respect to the fluid domain (i.e. pointing in towards the particle centre).

The development of the migration model in \citet{HardingStokes2020} may be summarised as the following sequence of steps:
\begin{enumerate}[label=\bfseries\arabic*.\, , leftmargin=\parindent]
\item
Introduce a rotating reference frame in which the particle's angular coordinate is constant. 
This frame rotates with angular velocity $\boldsymbol{\Theta}=\Theta\vec{k}$ where $\Theta:=\partial\theta_{p}/\partial t$.
Coordinates in this rotating frame are mapped to the `stationary frame' via
\begin{subequations}\begin{align}
\mathbf{x}'(r',\theta',z') &= (R+r')\cos(\theta')\mathbf{i}'+(R+r')\sin(\theta')\mathbf{j}'+z'\mathbf{k} \\
&= (R+r')\cos(\theta'+\theta_{p})\mathbf{i}+(R+r')\sin(\theta'+\theta_{p})\mathbf{j}+z'\mathbf{k} \,,
\end{align}\end{subequations}
where $\mathbf{i}':=\cos(\theta_{p})\mathbf{i}+\sin(\theta_{p})\mathbf{j}$ and $\mathbf{j}':=-\sin(\theta_{p})\mathbf{i}+\cos(\theta_{p})\mathbf{j}$.
It follows that in the rotating frame $\mathbf{x}_{p}^{\prime}=\mathbf{x}'(r_{p},0,z_{p})$, $\mathbf{u}'=\mathbf{u}-\boldsymbol{\Theta}\cross\mathbf{x}$, $\bar{\mathbf{u}}'=\bar{\mathbf{u}}-\boldsymbol{\Theta}\cross\mathbf{x}$, and so on.
\item
Assume the cross-sectional components of $\mathbf{u}_{p}$ are sufficiently small that the flow in the rotating frame is approximately stationary. 
As we are concerned with cross-sectional migration after any initial acceleration to terminal velocity in the axial direction, then acceleration effects (e.g. added mass and Basset force) may be neglected.
\item
Introduce the disturbance flow $q',\mathbf{v}'$ which satisfies
\begin{equation}
p'=\bar{p}'+q' \,, \qquad \vec{u}'=\bar{\vec{u}}'+\vec{v}' \,.
\end{equation}
\item
Non-dimensionalise using the velocity scale $U_{s}=(\alpha/2) U_{m}$ and length scale $a$, where $\alpha:=2a/\ell$. 
Most other scales may be derived from these (as per usual for a viscous flow).
The resulting Reynolds number $\Rey_{p}=(\rho/\mu)U_{s}a$ is often called the particle Reynolds number.
The force on the particle is non-dimensionalised via the scale $\Rey_{p}\mu U_{s}a$. 
Hats are used to describe non-dimensionalised variables, for example $\vec{v}'=U_{s}\hat{\vec{v}}'$.
\item
Apply a perturbation expansion to the disturbance flow with respect to $\Rey_{p}$, that is
\begin{equation}
\hat{\vec{v}}' = \vec{v}_{0} + \Rey_{p}\vec{v}_{1} + O(\Rey_{p}^{2}) \,, \qquad
\hat{q}' =  q_{0} + \Rey_{p} q_{1} + O(\Rey_{p}^{2}) \,.
\end{equation}
The force on the particle is expanded in a similar fashion, although the leading term has order $\Rey_{p}^{-1}$ and is denoted $\mathbf{F}_{-1}$ to reflect this, that is
\begin{equation}
\hat{\vec{F}}' = \Rey_{p}^{-1}\vec{F}_{-1} + \vec{F}_{0} + O(\Rey_{p}) \,,
\end{equation}
and similarly for the torque.
Observe we drop both the hat and prime from variables upon applying the perturbation expansion to each of $\hat{q}',\hat{\vec{v}}',\hat{\vec{F}}',\hat{\vec{T}}'$.
\end{enumerate}

Following these steps one arrives at: the equations 
\begin{subequations}\label{eqn:vq0}\begin{align}
\nabla\cdot\sigma(q_{0},\vec{v}_{0}) &= \mathbf{0} &&\text{for $\hat{\vec{x}}'\in\hat{\mathcal{F}}'$,} \\
\nabla\cdot\vec{v}_{0} &= 0 &&\text{for $\hat{\vec{x}}'\in\hat{\mathcal{F}}'$,} \\
\vec{v}_{0} &= \vec{0} &&\text{for $\hat{\vec{x}}'\in\partial\hat{\mathcal{D}}'$,} \\ 
\vec{v}_{0} &= \hat{\vec{u}}_{p,0}+\hat{\bs{\Omega}}_{p,0}\cross(\hat{\vec{x}}'-\hat{\vec{x}}_{p}^{\prime})-\hat{\bar{\mathbf{u}}} &&\text{for $\hat{\vec{x}}'\in\partial\hat{\mathcal{P}}'$,} \label{eqn:vq0d}
\end{align}\end{subequations}
governing $q_{0},\mathbf{v}_{0}$; the equations
\begin{subequations}\label{eqn:vq1}\begin{align}
\nabla\cdot\sigma(q_{1},\vec{v}_{1}) &= \hat{\boldsymbol{\Theta}}_0\cross\mathbf{v}_{0}+\mathbf{v}_{0}\cdot\nabla\hat{\bar{\mathbf{u}}}+(\mathbf{v}_{0}+\hat{\bar{\mathbf{u}}}-\hat{\boldsymbol{\Theta}}_0\cross\hat{\mathbf{x}}')\cdot\nabla\mathbf{v}_{0} &&\text{for $\hat{\vec{x}}'\in\hat{\mathcal{F}}'$,} \label{eqn:vq1_a}\\
\nabla\cdot\vec{v}_{1} &= 0 &&\text{for $\hat{\vec{x}}'\in\hat{\mathcal{F}}'$,} \\
\vec{v}_{1} &= \vec{0} &&\text{for $\hat{\vec{x}}'\in\partial\hat{\mathcal{D}}'$,} \\
\vec{v}_{1} &= \hat{\vec{u}}_{p,1}+\hat{\bs{\Omega}}_{p,1}\cross(\hat{\vec{x}}'-\hat{\vec{x}}_{p}^{\prime}) &&\text{for $\hat{\vec{x}}'\in\partial\hat{\mathcal{P}}'$,} \label{eqn:vq1d}
\end{align}\end{subequations}
governing $q_{1},\mathbf{v}_{1}$; and lastly, the force and torque terms of principle interest
\begin{subequations}\label{eqn:FT_expansion}\begin{align}
\mathbf{F}_{-1} &= \int_{\partial\hat{\mathcal{P}}'}(-\mathbf{n})\cdot\sigma(q_{0},\mathbf{v}_{0}) \,dS \,, \label{eqn:FL}\\
\mathbf{T}_{-1} &= \int_{\partial\hat{\mathcal{P}}'}(\hat{\mathbf{x}}'-\hat{\mathbf{x}}_{p}^{\prime})\cross\left((-\mathbf{n})\cdot\sigma(q_{0},\mathbf{v}_{0})\right) \,dS \,, \label{eqn:TL} \\
\mathbf{F}_{0} &= -\frac{4\pi}{3}\hat{\boldsymbol{\Theta}}_{0}\cross(\hat{\boldsymbol{\Theta}}_{0}\cross\hat{\mathbf{x}}_{p}^{\prime})
-\frac{8\pi}{3}\hat{\boldsymbol{\Theta}}_{0}\cross\hat{\vec{u}}_{p,0}^\prime 
+\int_{\hat{\mathcal{P}}'}\hat{\bar{\mathbf{u}}}\cdot\nabla\hat{\bar{\mathbf{u}}}\,dV \notag\\&\hspace{2.6cm}
+\int_{\partial\hat{\mathcal{P}}'}(-\mathbf{n})\cdot\sigma(q_{1},\mathbf{v}_{1}) \,dS \,, \label{eqn:F0} \\
\mathbf{T}_{0} &= 
-\frac{8\pi}{15}\hat{\boldsymbol{\Theta}}_{0}\cross\hat{\boldsymbol{\Omega}}_{p,0}
+\int_{\hat{\mathcal{P}}'}(\hat{\mathbf{x}}'-\hat{\mathbf{x}}_{p}^{\prime})\cross(\hat{\bar{\mathbf{u}}}\cdot\nabla\hat{\bar{\mathbf{u}}})\,dV \notag\\&\hspace{2.6cm}
+\int_{\partial\hat{\mathcal{P}}'}(\hat{\mathbf{x}}'-\hat{\mathbf{x}}_{p}^{\prime})\cross\left[(-\mathbf{n})\cdot\sigma(q_{1},\mathbf{v}_{1})\right]\,dS \,. \label{eqn:T0}
\end{align}\end{subequations}

We refer the interested reader to \citet{HardingStokesBertozzi2019} for a complete derivation.
Note that in \eqref{eqn:vq0d} we retain a $\hat{\vec{u}}_{p}$ contribution whereas in \citet{HardingStokesBertozzi2019} this was taken to be $\bs{\Theta}\times\mathbf{x}_p^\prime$ (i.e. the $\theta$ component only) with drag coefficients associated with the cross-sectional components introduced later.
Additionally, here we include particle velocity and spin contributions in \eqref{eqn:vq1d}.
Notice these contributions in both \eqref{eqn:vq0d} and \eqref{eqn:vq1d} are expressed in terms of $\hat{\vec{u}}_{p},\hat{\bs{\Omega}}_{p}$ as viewed in the stationary reference frame.
These minor changes allow us to separate the leading and first order contributions to the particle motion, and account for inertial lift contributions arising from the secondary component of the background flow.

It is reasonably well-known that the last term of \eqref{eqn:F0} can be computed without needing to explicitly calculate $q_1,\mathbf{v}_1$ via an application of the Lorenz reciprocal theorem.
Thus we only need to consider the solution of \eqref{eqn:vq0}.
Before moving on to describe our improved background flow approximation, we introduce some additional notation and make a remark about the use of the Lorentz reciprocal theorem.

\subsection{Stokes' flow operators}\label{sec:stokes_operators}

We introduce the operators $\mathcal{Q}(\mathbf{f}),\boldsymbol{\mathcal{V}}(\mathbf{f})$ which map a continuous vector field $\vec{f}$ defined on $\partial\hat{\mathcal{P}}'$ to a pressure and velocity field, respectively, which satisfies the Stokes' equations
\begin{subequations}\label{eqn:vq_ast}\begin{align}
\nabla\cdot\sigma\left(\mathcal{Q}(\mathbf{f}),\boldsymbol{\mathcal{V}}(\mathbf{f})\right) &= \mathbf{0} &&\text{for $\hat{\vec{x}}'\in\hat{\mathcal{F}}'$,} \\
\nabla\cdot\boldsymbol{\mathcal{V}}(\mathbf{f}) &= 0 &&\text{for $\hat{\vec{x}}'\in\hat{\mathcal{F}}'$,} \\
\boldsymbol{\mathcal{V}}(\mathbf{f}) &= \mathbf{0} &&\text{for $\hat{\vec{x}}'\in\partial\hat{\mathcal{D}}'$,} \\ 
\boldsymbol{\mathcal{V}}(\mathbf{f}) &= \mathbf{f} &&\text{for $\hat{\vec{x}}'\in\partial\hat{\mathcal{P}}'$.} 
\end{align}\end{subequations}
Additionally, the corresponding hydrodynamic force and torque on the particle are denoted by
\begin{subequations}\label{eqn:FT_coef}\begin{align}
\boldsymbol{\mathcal{M}}(\mathbf{f}) &:= \int_{\partial\hat{\mathcal{P}}'} (-\mathbf{n})\cdot\sigma(\mathcal{Q}(\mathbf{f}),\boldsymbol{\mathcal{V}}(\mathbf{f})) \,dS \,, \\
\boldsymbol{\mathcal{N}}(\mathbf{f}) &:= \int_{\partial\hat{\mathcal{P}}'} (\hat{\mathbf{x}}'-\hat{\mathbf{x}}_{p}^{\prime})\cross\left((-\mathbf{n})\cdot\sigma(\mathcal{Q}(\mathbf{f}),\boldsymbol{\mathcal{V}}(\mathbf{f}))\right) \,dS \,. 
\end{align}\end{subequations}
When $\mathbf{f}$ is a constant unit vector then $\boldsymbol{\mathcal{M}}(\mathbf{f})$ and $\boldsymbol{\mathcal{N}}(\mathbf{f})$ may be interpreted as force and torque coefficients, respectively, in the direction of $\mathbf{f}$.
Note that all four of these operators implicitly depend on the location of the particle within the cross-section since the fluid domain $\hat{\mathcal{F}}'$ and particle boundary $\partial\hat{\mathcal{P}}'$ depend on $\hat{r}_p,\hat{z}_p$.
Additionally, there is an implicit dependence on $\epsilon$, as this relates to the bend radius of the duct, and $\alpha$, as this relates to the size of the non-dimensionalised fluid domain. 
The implicit dependence on $\epsilon$ is expected to be very weak for $\epsilon\leq 1/100$.

Observe that the operators $\mathcal{Q},\boldsymbol{\mathcal{V}},\boldsymbol{\mathcal{M}},\boldsymbol{\mathcal{N}}$ are linear and therefore, for example, we may write 
\begin{subequations}\label{eqn:linearity}\begin{align}
q_{0} &=\mathcal{Q}(\hat{\vec{u}}_{p,0})+\mathcal{Q}(\hat{\bs{\Omega}}_{p,0}\cross(\hat{\vec{x}}'-\hat{\vec{x}}_{p}^{\prime}))-\mathcal{Q}(\hat{\bar{\mathbf{u}}}) \,, \\
\mathbf{v}_{0} &=\boldsymbol{\mathcal{V}}(\hat{\vec{u}}_{p,0})+\boldsymbol{\mathcal{V}}(\hat{\bs{\Omega}}_{p,0}\cross(\hat{\vec{x}}'-\hat{\vec{x}}_{p}^{\prime}))-\boldsymbol{\mathcal{V}}(\hat{\bar{\mathbf{u}}}) \,, \\
\mathbf{F}_{-1} &=\boldsymbol{\mathcal{M}}(\hat{\vec{u}}_{p,0})+\boldsymbol{\mathcal{M}}(\hat{\bs{\Omega}}_{p,0}\cross(\hat{\vec{x}}'-\hat{\vec{x}}_{p}^{\prime}))-\boldsymbol{\mathcal{M}}(\hat{\bar{\mathbf{u}}}) \,, \\
\mathbf{T}_{-1} &=\boldsymbol{\mathcal{N}}(\hat{\vec{u}}_{p,0})+\boldsymbol{\mathcal{N}}(\hat{\bs{\Omega}}_{p,0}\cross(\hat{\vec{x}}'-\hat{\vec{x}}_{p}^{\prime}))-\boldsymbol{\mathcal{N}}(\hat{\bar{\mathbf{u}}}) \,.
\end{align}\end{subequations}
This linearity will be further exploited in Section~\ref{sec:qv0_expansion}.
Appendix~\ref{app:symmetries} describes when various components of these coefficients are zero due to the symmetry of our domain (in the $\theta$ direction about $\theta_p$).

\subsection{A note on reciprocal theorems}

A variant of the Lorentz reciprocal theorem can be applied to show that 
\begin{multline}\label{eqn:force_recip}
\int_{\partial\hat{\mathcal{P}}'}(-\mathbf{n})\cdot\sigma(q_{1},\mathbf{v}_{1}) \,dS 
= \bs{\mathcal{M}}(\hat{\vec{u}}_{p,1})+\bs{\mathcal{M}}(\hat{\bs{\Omega}}_{p,1}\cross(\hat{\vec{x}}'-\hat{\vec{x}}_{p}^{\prime})) \\
-\sum_{\vec{e}\in\{\vec{i}',\,\vec{j}',\,\vec{k}\}}\vec{e}\int_{\hat{\mathcal{F}}'}\bs{\mathcal{V}}(\vec{e})\cdot\mathbf{I}_{0} \,dV \,,
\end{multline}
where $\mathbf{I}_{0}$ is the right side of \eqref{eqn:vq1_a}, that is 
\begin{equation}
\mathbf{I}_{0} := \hat{\bs{\Theta}}_{0}\cross\mathbf{v}_{0}+\mathbf{v}_{0}\cdot\nabla\hat{\bar{\mathbf{u}}}+(\mathbf{v}_{0}+\hat{\bar{\mathbf{u}}}-\hat{\bs{\Theta}}_{0}\cross\hat{\mathbf{x}}')\cdot\nabla\mathbf{v}_{0} \,.
\end{equation}
This application of the Lorentz reciprocal theorem to the calculation of the inertial lift force is reasonably well-known. 
Perhaps less well-known is that it can also be applied to the calculation of the torque terms.
Specifically, the third term of $\mathbf{T}_{0}$ in \eqref{eqn:T0} can be calculated via
\begin{multline}\label{eqn:torque_recip}
\int_{\partial\hat{\mathcal{P}}'}(\hat{\vec{x}}'-\hat{\vec{x}}_{p}^{\prime})\cross\left(-\mathbf{n}\cdot\sigma(q_{1},\mathbf{v}_{1})\right) \,dS 
= \bs{\mathcal{N}}(\hat{\vec{u}}_{p,1}) +\bs{\mathcal{N}}(\hat{\bs{\Omega}}_{p,1}\cross(\hat{\vec{x}}'-\hat{\vec{x}}_{p}^{\prime})) \\
-\sum_{\vec{e}\in\{\vec{i}',\,\vec{j}',\,\vec{k}\}}\vec{e}\int_{\hat{\mathcal{F}}'}\boldsymbol{\mathcal{V}}(\vec{e}\cross(\hat{\vec{x}}'-\hat{\vec{x}}_{p}^{\prime}))\cdot\mathbf{I}_{0} \,dV \,.
\end{multline}
Notice that the $\bs{\mathcal{M}}(\cdot)$ and $\bs{\mathcal{N}}(\cdot)$ terms in \eqref{eqn:force_recip} and \eqref{eqn:torque_recip}, respectively, encapsulate the contributions from the the boundary conditions \eqref{eqn:vq1d}, while the remaining volume integrals are the result of the reciprocal theorem applied to capture the contribution of $\mathbf{I}_{0}$.
A detailed proof of the reciprocal theorems that give rise to these volume integrals is provided in Appendix~\ref{app:recip}.

\section{Improved approximation of the background flow}\label{sec:bg_flow}

The background flow $\bar{p},\bar{\mathbf{u}}$ is a steady solution of the Navier--Stokes equations:
\begin{subequations}\label{eqn:NS_bg}\begin{align}
\nabla\cdot\sigma(\bar{p},\mathbf{\bar{u}}) &= \rho\,\bar{\mathbf{u}}\cdot\nabla\bar{\mathbf{u}} && \mathbf{x}\in\mathcal{D}, \\
\nabla\cdot\bar{\mathbf{u}} &= 0 && \mathbf{x}\in\mathcal{D}, \\
\bar{\mathbf{u}} &= \mathbf{0} && \mathbf{x}\in\partial\mathcal{D}.
\end{align}\end{subequations}
The resulting velocity field may be decomposed into its axial component $\bar{\mathbf{u}}_a$ and secondary component $\bar{\mathbf{u}}_s$.
A perturbation expansion may be applied to each component with respect to $K=\epsilon\Rey^2$ which converges provided $K\lesssim 212$ \citep{HardingANZIAMJ2019}.
Specifically
\begin{subequations}\label{eqn:bgpa}
\begin{align}
\bar{\mathbf{u}}_a &= U_m\sum_{i=0}^{\infty}K^i\bar{u}_{\theta,i}\mathbf{e}_{\theta} \,, \\
\bar{\mathbf{u}}_s &= U_m\epsilon\Rey\sum_{i=0}^{\infty}K^i(\bar{u}_{r,i}\mathbf{e}_{r}+\bar{u}_{z,i}\mathbf{e}_{z}) \,,\label{eqn:bgpa_b}
\end{align}
\end{subequations}
where $\bar{u}_{\theta,i},\bar{u}_{r,i},\bar{u}_{z,i}$ are the (dimensionless) components of $\bar{\vec{u}}$ in the $\theta,r,z$ directions, respectively, and are each independent of $\theta$.

Our previous model used only the leading order terms $\bar{u}_{\theta,0},\bar{u}_{r,0},\bar{u}_{z,0}$ to model inertial migration in curved ducts based on an assumption that $K$ is suitably small \citep{HardingStokesBertozzi2019}.
In this study we extend the use of \eqref{eqn:bgpa} to accurately model particle migration for values of $K$ up to $O(100)$.
Specifically, we will use the three leading terms $i=0,1,2$ to construct a model of particle migration which is quadratic in $K$.
Upon non-dimensionalising with respect to the shear velocity scale ($U_s=(\alpha/2) U_m=(a/\ell)U_m$) our model may be expressed as
\begin{equation}\label{eqn:ubar_pert}
\hat{\bar{\mathbf{u}}}\approx 2\alpha^{-1}\left(\bar{\vec{u}}_{a,0}+K\bar{\vec{u}}_{a,1}+K^{2}\bar{\vec{u}}_{a,2}\right)
+\kappa\Rey_{p}\left(\bar{\mathbf{u}}_{s,0}+K\bar{\mathbf{u}}_{s,1}+K^2\bar{\mathbf{u}}_{s,2}\right) \,,
\end{equation}
where $\bar{\mathbf{u}}_{a,i}:=\bar{u}_{\theta,i}\mathbf{e}_{\theta}$ and $\bar{\mathbf{u}}_{s,i}:=\bar{u}_{r,i}\mathbf{e}_{r}+\bar{u}_{z,i}\mathbf{e}_{z}$ for each $i$.
Notice in \eqref{eqn:ubar_pert} we have used the fact that $2\alpha^{-1}\epsilon\Rey=\kappa\Rey_p$ (recalling $\kappa=\ell^{4}/(4a^3R)$).

\begin{figure}
\centering
\begin{subfigure}[b]{0.54\textwidth} 
\centering
\includegraphics{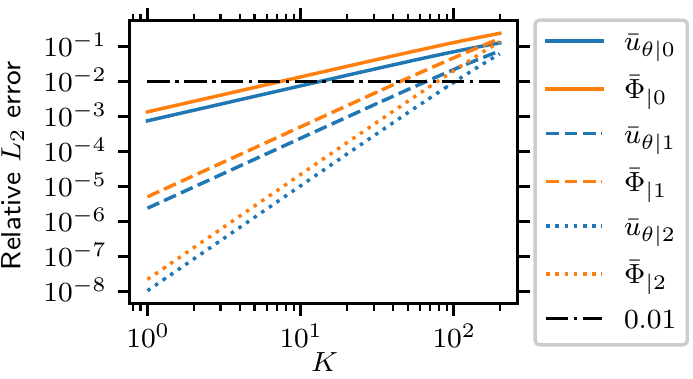}
\caption{$W/H=2$ and $\epsilon=0.01$}\label{fig:uPhi_trends_a}
\end{subfigure}
\begin{subfigure}[b]{0.43\textwidth}
\centering
\includegraphics{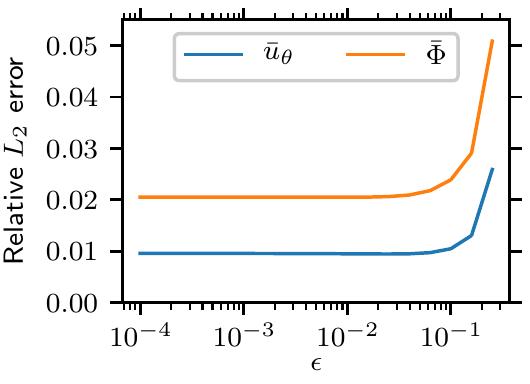}
\caption{$W/H=2$ and $K=100$}\label{fig:uPhi_trends_b}
\end{subfigure}
\\
\begin{subfigure}[b]{0.43\textwidth}
\centering
\includegraphics{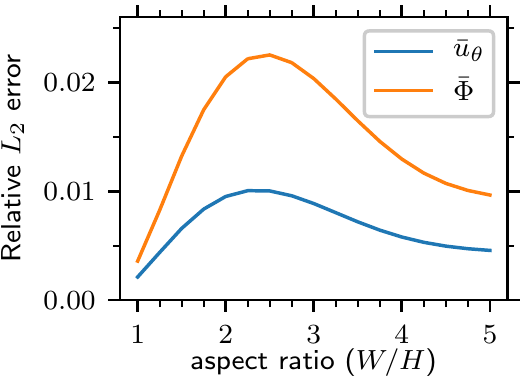}
\caption{$K=100$ and $\epsilon=0.01$}\label{fig:uPhi_trends_c}
\end{subfigure}
\begin{subfigure}[b]{0.54\textwidth}
\centering
\includegraphics{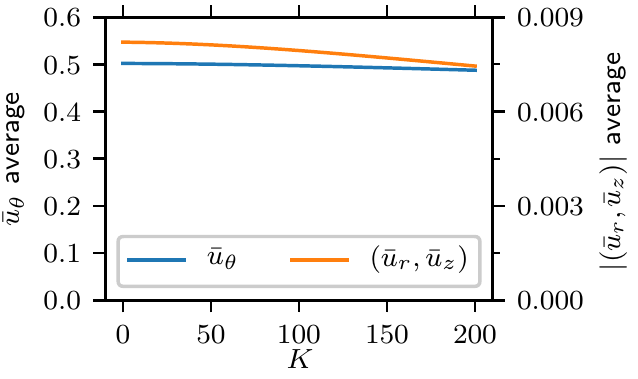}
\caption{$W/H=2$ and $\epsilon=0.01$}\label{fig:uPhi_trends_d}
\end{subfigure}
\caption{
The relative $L_{2}$ error of truncated perturbation approximations of $\bar{u}_{\theta}$ and $\bar{\Phi}$ versus (a) the Dean number $K\in[1,200]$, (b) the relative curvature $\epsilon\in[10^{-4},0.25]$, and (c) the cross-section aspect ratio $W/H\in[1,5]$. The last plot (d) shows the change in the average of $\bar{u}_{\theta}$ and $|(\bar{u}_{r},\bar{u}_{z})|$ versus $K\in[0,200]$.}\label{fig:uPhi_trends}
\end{figure}

We illustrate the accuracy of our improved model of the background flow in Figure~\ref{fig:uPhi_trends}.
We use the streamfunction $\bar{\Phi}$, for which $\bar{u}_{r}=-(1+\epsilon r)^{-1}\partial\bar{\Phi}/\partial z$ and $\bar{u}_{z}=(1+\epsilon r)^{-1}\partial\bar{\Phi}/\partial r$, 
for the purpose of evaluating the accuracy of the secondary components.
Figure~\ref{fig:uPhi_trends_a} illustrates the accuracy of our quadratic model over the range $1\leq K\leq 200$ using a fixed aspect ratio $W/H=2$ and curvature parameter $\epsilon=0.01$.
Here, the notation $\bar{u}_{\theta\mid i},\bar{\Phi}_{\mid i}$ indicates the approximation \eqref{eqn:bgpa} is truncated beyond the $i$'th term.
In each case the relative $L_2$ error is proportional to $K^{i+1}$. 
Observe that the approximations $\bar{u}_{\theta\mid 1},\bar{\Phi}_{\mid 1}$ and $\bar{u}_{\theta\mid 2},\bar{\Phi}_{\mid 2}$ are a significant improvement over $\bar{u}_{\theta\mid 0},\bar{\Phi}_{\mid 0}$ for $K=O(10)$.
Additionally, for $K=O(100)$ the approximation $\bar{u}_{\theta\mid 2},\bar{\Phi}_{\mid 2}$ offers a marginal improvement over $\bar{u}_{\theta\mid 1},\bar{\Phi}_{\mid 1}$.
When $K=100$ the relative errors are $0.95\%$ for $\bar{u}_{\theta\mid2}$ and $2.05\%$ for $\bar{\Phi}_{\mid2}$.
This illustrates that the first three terms is a reasonable trade-off between computational cost and achieving a reasonably accurate model for $K=O(100)$.

For the remainder of the discussion we take $\bar{u}_{\theta}=\bar{u}_{\theta\mid 2}$ and similarly $\bar{\Phi}=\bar{\Phi}_{\mid2}$.
Figure~\ref{fig:uPhi_trends_b} illustrates that the model accuracy is insensitive to bend radius when $\epsilon<1/10$, using fixed $K=100$ and aspect ratio $W/H=2$.
Figure~\ref{fig:uPhi_trends_c} shows the variation in model error with respect to aspect ratio of the cross-section, using fixed $K=100$ and $\epsilon=0.01$. 
While there is some dependence of the model error on the aspect ratio, the case $W/H=2$ is near the peak for aspect ratios $W/H\geq1$ and is therefore a reasonable choice of reference value.

We take a moment to elaborate on the choice of characteristic velocity $U_m$ used in this study.
The pressure gradient which drives flow through the curved duct increases super-linearly with the flow rate and also has a (weak) dependence on the bend radius.
Given a $K>0$, determining the pressure gradient which produces the required flow rate becomes a non-linear optimisation problem.
The non-linearity is weak over the range of $K$ considered in this study and therefore it will be convenient to avoid the non-linear optimisation.
To this end, we specify the characteristic velocity $U_m$ to be the maximum velocity of a laminar flow through a straight duct having an identical cross-section. 
This is equivalent to non-dimensionalising with respect to the driving pressure gradient, since straight duct flow satisfies Stokes' equation and therefore doesn't depend on the Reynolds number, that is $U_m=c_{\mathcal{C}}G\ell^2/\mu$ where $G$ is the pressure gradient down the main flow axis and $c_{\mathcal{C}}$ is a constant depending only on the cross-section $\mathcal{C}$.
This choice makes it straightforward to compare results across different duct bend radii and incorporate $K$ as a new parameter in our inertial migration model.
Figure~\ref{fig:uPhi_trends_d} illustrates how the average of the non-dimensional $\bar{u}_\theta$ and $\bar{\Phi}$ varies as a function of $K$ for our specific non-dimensionalisation, with fixed $\epsilon=0.01$ and $W/H=2$. 
The average of $\bar{u}_\theta$ is approximately constant over this range of $K$ indicating that our specific choice of characteristic velocity is easily translated to an average axial flow rate if desired.

\begin{figure}
\centering
\begin{subfigure}[b]{0.48\textwidth}
\centering
\includegraphics{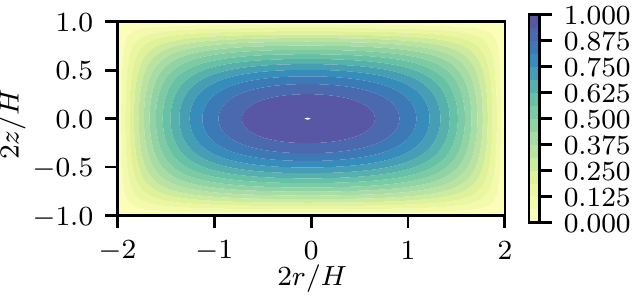}
\caption{$\bar{u}_\theta$ with $K=0$}
\end{subfigure}
\begin{subfigure}[b]{0.51\textwidth}
\centering
\includegraphics{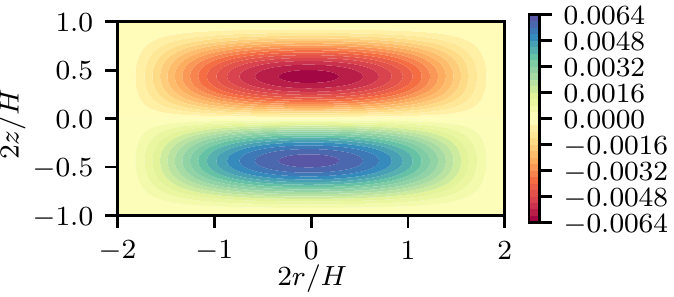}
\caption{$\bar{\Phi}$ with $K=0$}
\end{subfigure}
\\
\begin{subfigure}[b]{0.48\textwidth}
\centering
\includegraphics{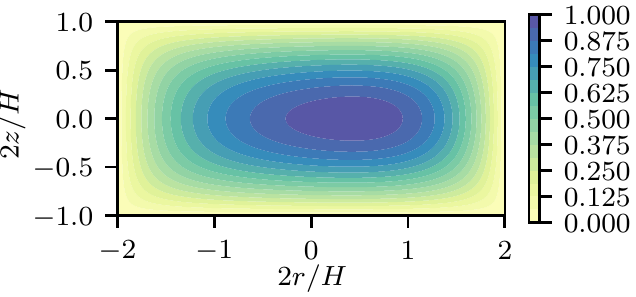}
\caption{$\bar{u}_\theta$ with $K=100$}
\end{subfigure}
\begin{subfigure}[b]{0.51\textwidth}
\centering
\includegraphics{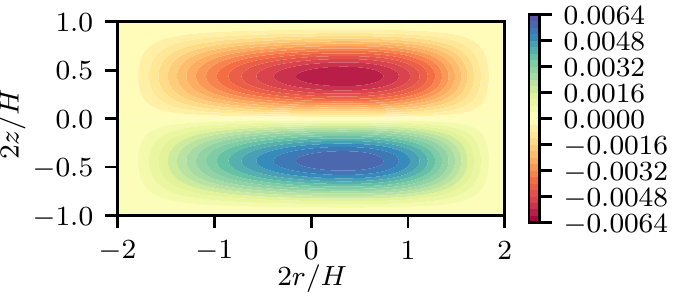}
\caption{$\bar{\Phi}$ with $K=100$}
\end{subfigure}
\caption{
The fields (a,c) $\bar{u}_\theta$ and (b,d) $\bar{\Phi}$ for (a,b) $K=0$ and (c,d) $K=100$.
In each case $\epsilon=0.01$ and $W/H=2$.
The colour bars have been fixed across the pairs (a,c) and (b,d) for comparison.}\label{fig:uPhi_plots}
\end{figure}

Figure~\ref{fig:uPhi_plots} illustrates the difference in features of the fields $\bar{u}_\theta,\bar{\Phi}$ between $K=0$ to $K=100$. 
Note that $K=0$ should be interpreted as the limit $K\to0$ as a result of a decaying flow rate. $U_m$ 
While the physical magnitude of the secondary vortices decays to zero as $U_m\to0$ our specific non-dimensionalisation of $\bar{\mathbf{u}}_s$, via the scaling $U_m\epsilon\Rey$ in \eqref{eqn:bgpa_b}, ensures that the magnitude of the secondary vortices remains finite.
Observe in Figure~\ref{fig:uPhi_plots} a shift of local maxima and minima towards the outside/right wall when $K=100$. 
This is due to the increased rotational inertia of the fluid and is not captured by the leading order approximation of the background flow used in \citet{HardingStokesBertozzi2019}.
This shift becomes more pronounced with further increases in $K$. 
Lastly, we point out that the Dean numbers $K\lesssim 200$ considered herein are significantly smaller than the critical Dean number after which there exist four vortex solutions~\citep{Winters1987}.

\section{Our extended particle migration model}\label{sec:qv0_expansion}

Here we incorporate the improved approximation of the background flow \eqref{eqn:ubar_pert} into the inertial lift model so that we may efficiently approximate particle trajectories for any desired $0\leq K\lesssim200$.
We start by decomposing the leading order disturbance flow $q_{0},\mathbf{v}_{0}$ further than was previously done in \citet{HardingStokesBertozzi2019}.
This approach has the additional benefit of revealing how the axial and secondary components of the background flow influence all six components of $\hat{\vec{u}}_{p},\hat{\bs{\Omega}}_{p}$ at both leading and first order. 
Additionally, the decomposition admits a decoupling which can be exploited to more efficiently compute all of the required forces and torques.

Utilising the operator $\bs{\mathcal{V}}$, and expanding further on \eqref{eqn:linearity}, observe that $\mathbf{v}_{0}$ can be decomposed as
\begin{align}
\mathbf{v}_{0} &= u_{p,0,r}\bs{\mathcal{V}}(\mathbf{i}')
+u_{p,0,\theta}\bs{\mathcal{V}}(\mathbf{j}')
+u_{p,0,z}\bs{\mathcal{V}}(\mathbf{k}) \notag\\
&\quad+\Omega_{p,0,r}\bs{\mathcal{V}}(\mathbf{i}'\cross(\hat{\vec{x}}'-\hat{\vec{x}}_{p}^{\prime}))+\Omega_{p,0,\theta}\bs{\mathcal{V}}(\mathbf{j}'\cross(\hat{\vec{x}}'-\hat{\vec{x}}_{p}^{\prime}))+\Omega_{p,0,z}\bs{\mathcal{V}}(\mathbf{k}\cross(\hat{\vec{x}}'-\hat{\vec{x}}_{p}^{\prime})) \notag\\
&\quad-2\alpha^{-1}
\left[\bs{\mathcal{V}}(\bar{\vec{u}}_{a,0})+K\bs{\mathcal{V}}(\bar{\vec{u}}_{a,1})+K^{2}\bs{\mathcal{V}}(\bar{\vec{u}}_{a,2})\right] \notag\\
&\quad-\kappa\Rey_{p}\left[\bs{\mathcal{V}}(\bar{\mathbf{u}}_{s,0}) +K\bs{\mathcal{V}}(\bar{\mathbf{u}}_{s,1})+K^{2}\bs{\mathcal{V}}(\bar{\mathbf{u}}_{s,2})\right]  \,, \label{eqn:v0_expansion}
\end{align}
where $\hat{\vec{u}}_{p,0}=(u_{p,0,r},u_{p,0,\theta},u_{p,0,z})$ and $\hat{\bs{\Omega}}_{p,0}=(\Omega_{p,0,r},\Omega_{p,0,\theta},\Omega_{p,0,z})$.
This expansion extracts all of the key parameters from the leading disturbance solution making it a linear superposition of Stokes' flow solutions.

Notice that last line of \eqref{eqn:v0_expansion} has a factor $\Rey_p$ and could therefore be placed in $\mathbf{v}_{1}$ (which would be equivalent to moving the appropriate component of the boundary condition in \eqref{eqn:vq0d} to \eqref{eqn:vq1d}). 
However, in many practical situations $\kappa\gg1$ and thus $\kappa\Rey_p$ may not be small.
Consequently, it will be convenient to keep these terms in $\vec{v}_0$. 
An expansion identical to \eqref{eqn:v0_expansion} is obtained for each of $q_{0}$, $\mathbf{F}_{-1}$ and $\mathbf{T}_{-1}$ by replacing each $\boldsymbol{\mathcal{V}}$ with $\mathcal{Q}$, $\boldsymbol{\mathcal{M}}$ and $\boldsymbol{\mathcal{N}}$, respectively.

Upon setting $\vec{F}_{-1}=\vec{0}$ and $\vec{T}_{-1}=\vec{0}$, one obtains six equations which are linear with respect to the six components of $\mathbf{u}_{p,0},\boldsymbol{\Omega}_{p,0}$.
Applying the symmetry properties described in Appendix~\ref{app:symmetries} allows the $6\cross6$ system to be decoupled into two distinct $3\cross3$ systems.
One of these describes the equilibrium attained by the particle parameters $u_{p,0,\theta},\Omega_{p,0,r},\Omega_{p,0,z}$ in relation to the axial motion of the background flow:
\begin{equation}\label{eqn:leading_vel_spin}
A_a \cdot \begin{bmatrix}
u_{p,0,\theta} \\
\Omega_{p,0,r} \\
\Omega_{p,0,z}
\end{bmatrix} 
=2\alpha^{-1}
\begin{bmatrix}
\left[\boldsymbol{\mathcal{M}}(\bar{\vec{u}}_{a,0})+K\boldsymbol{\mathcal{M}}(\bar{\vec{u}}_{a,1})+K^{2}\boldsymbol{\mathcal{M}}(\bar{\vec{u}}_{a,2})\right]\cdot\mathbf{j}' \\[2pt]
\left[\boldsymbol{\mathcal{N}}(\bar{\vec{u}}_{a,0})+K\boldsymbol{\mathcal{N}}(\bar{\vec{u}}_{a,1})+K^{2}\boldsymbol{\mathcal{N}}(\bar{\vec{u}}_{a,2})\right]\cdot\mathbf{i}' \\[2pt]
\left[\boldsymbol{\mathcal{N}}(\bar{\vec{u}}_{a,0})+K\boldsymbol{\mathcal{N}}(\bar{\vec{u}}_{a,1})+K^{2}\boldsymbol{\mathcal{N}}(\bar{\vec{u}}_{a,2})\right]\cdot\mathbf{k} 
\end{bmatrix} \,,
\end{equation}
where
\begin{equation}
A_a := \begin{bmatrix}
\boldsymbol{\mathcal{M}}(\mathbf{j}')\cdot\mathbf{j}' & \boldsymbol{\mathcal{M}}(\mathbf{i}'\cross(\hat{\vec{x}}'-\hat{\vec{x}}_{p}^{\prime}))\cdot\mathbf{j}' & \boldsymbol{\mathcal{M}}(\mathbf{k}\cross(\hat{\vec{x}}'-\hat{\vec{x}}_{p}^{\prime}))\cdot\mathbf{j}' \\
\boldsymbol{\mathcal{N}}(\mathbf{j}')\cdot\mathbf{i}' & \boldsymbol{\mathcal{N}}(\mathbf{i}'\cross(\hat{\vec{x}}'-\hat{\vec{x}}_{p}^{\prime}))\cdot\mathbf{i}' & \boldsymbol{\mathcal{N}}(\mathbf{k}\cross(\hat{\vec{x}}'-\hat{\vec{x}}_{p}^{\prime}))\cdot\mathbf{i}' \\
\boldsymbol{\mathcal{N}}(\mathbf{j}')\cdot\mathbf{k} & \boldsymbol{\mathcal{N}}(\mathbf{i}'\cross(\hat{\vec{x}}'-\hat{\vec{x}}_{p}^{\prime}))\cdot\mathbf{k} & \boldsymbol{\mathcal{N}}(\mathbf{k}\cross(\hat{\vec{x}}'-\hat{\vec{x}}_{p}^{\prime}))\cdot\mathbf{k}
\end{bmatrix}\,.
\end{equation}
The diagonal elements of $A_a$ are negative and generally the matrix is diagonally dominant. 
Recall that each $\boldsymbol{\mathcal{M}}(\cdot)$ and $\boldsymbol{\mathcal{N}}(\cdot)$ implicitly depends on $\epsilon,\alpha$ and the cross-sectional coordinates of the particle $(\hat{r}_{p},\hat{z}_{p})$.
Solving \eqref{eqn:leading_vel_spin} therefore yields an expression for each of $u_{p,0,\theta},\Omega_{p,0,r},\Omega_{p,0,z}$ which depend only on the parameters $\hat{r}_{p},\hat{z}_{p},\epsilon,\alpha,K$.

The remaining $3\cross3$ system describes the equilibrium attained by the particle parameters $u_{p,r},u_{p,z},\Omega_{p,\theta}$ in relation to the secondary motion of the background flow:
\begin{equation}\label{eqn:cs_vel_spin}
A_s \cdot \begin{bmatrix}
u_{p,0,r} \\
u_{p,0,z} \\
\Omega_{p,0,\theta}
\end{bmatrix}
=\kappa\Rey_{p}\begin{bmatrix}
\left[\boldsymbol{\mathcal{M}}(\bar{\mathbf{u}}_{s,0})+K\boldsymbol{\mathcal{M}}(\bar{\mathbf{u}}_{s,1})+K^{2}\boldsymbol{\mathcal{M}}(\bar{\mathbf{u}}_{s,2})\right]\cdot\mathbf{i}' \\[2pt]
\left[\boldsymbol{\mathcal{M}}(\bar{\mathbf{u}}_{s,0})+K\boldsymbol{\mathcal{M}}(\bar{\mathbf{u}}_{s,1})+K^{2}\boldsymbol{\mathcal{M}}(\bar{\mathbf{u}}_{s,2})\right]\cdot\mathbf{k} \\[2pt]
\left[\boldsymbol{\mathcal{N}}(\bar{\mathbf{u}}_{s,0})+K\boldsymbol{\mathcal{N}}(\bar{\mathbf{u}}_{s,1})+K^{2}\boldsymbol{\mathcal{N}}(\bar{\mathbf{u}}_{s,2})\right]\cdot\mathbf{j}' 
\end{bmatrix} \,,
\end{equation}
where
\begin{equation}
A_s := \begin{bmatrix}
\boldsymbol{\mathcal{M}}(\mathbf{i}')\cdot\mathbf{i}' & \boldsymbol{\mathcal{M}}(\mathbf{k})\cdot\mathbf{i}' & \boldsymbol{\mathcal{M}}(\mathbf{j}'\cross(\hat{\vec{x}}'-\hat{\vec{x}}_{p}^{\prime}))\cdot\mathbf{i}' \\
\boldsymbol{\mathcal{M}}(\mathbf{i}')\cdot\mathbf{k} & \boldsymbol{\mathcal{M}}(\mathbf{k})\cdot\mathbf{k} & \boldsymbol{\mathcal{M}}(\mathbf{j}'\cross(\hat{\vec{x}}'-\hat{\vec{x}}_{p}^{\prime}))\cdot\mathbf{k} \\
\boldsymbol{\mathcal{N}}(\mathbf{i}')\cdot\mathbf{j}' & \boldsymbol{\mathcal{N}}(\mathbf{k})\cdot\mathbf{j}' & \boldsymbol{\mathcal{N}}(\mathbf{j}'\cross(\hat{\vec{x}}'-\hat{\vec{x}}_{p}^{\prime}))\cdot\mathbf{j}' 
\end{bmatrix} \,.
\end{equation}
Similar to $A_a$, the diagonal elements of $A_s$ are negative and generally the matrix is diagonally dominant.

We now consider the balance of force and torque at the next order of $\Rey_p$ by setting $\vec{F}_{0}=\vec{0}$ and $\vec{T}_{0}=\vec{0}$.
The symmetries that lead to \eqref{eqn:leading_vel_spin} and \eqref{eqn:cs_vel_spin} analogously lead to
\begin{multline}\label{eqn:a1_vel_spin}
A_a
\begin{bmatrix}
u_{p,1,\theta} \\
\Omega_{p,1,r} \\
\Omega_{p,1,z}
\end{bmatrix} 
=
\begin{bmatrix}
\frac{8\pi}{3}\Theta_0 u_{p,0,r} \\[2pt]  
-\frac{8\pi}{15}\Theta_{0}\Omega_{p,0,\theta} \\[2pt] 
0 
\end{bmatrix}
-\begin{bmatrix}
\vec{j}'\cdot\int_{\hat{\mathcal{P}}'}\hat{\bar{\vec{u}}}\cdot\nabla\hat{\bar{\vec{u}}}\,dV \\[2pt]
\vec{i}'\cdot\int_{\hat{\mathcal{P}}'}(\hat{\vec{x}}'-\hat{\vec{x}}_{p}^{\prime})\cross(\hat{\bar{\vec{u}}}\cdot\nabla\hat{\bar{\vec{u}}})\,dV \\[2pt]
\vec{k}\cdot\int_{\hat{\mathcal{P}}'}(\hat{\vec{x}}'-\hat{\vec{x}}_{p}^{\prime})\cross(\hat{\bar{\vec{u}}}\cdot\nabla\hat{\bar{\vec{u}}})\,dV
\end{bmatrix} \\
+\begin{bmatrix}	
\int_{\hat{\mathcal{F}}'}\boldsymbol{\mathcal{V}}(\vec{j}')\cdot\mathbf{I}_{0} \,dV \\[2pt]
\int_{\hat{\mathcal{F}}'}\boldsymbol{\mathcal{V}}(\vec{i}'\cross(\hat{\vec{x}}'-\hat{\vec{x}}_{p}^{\prime}))\cdot\mathbf{I}_{0} \,dV \\[2pt]
\int_{\hat{\mathcal{F}}'}\boldsymbol{\mathcal{V}}(\vec{k}\cross(\hat{\vec{x}}'-\hat{\vec{x}}_{p}^{\prime}))\cdot\mathbf{I}_{0} \,dV
\end{bmatrix} \,,
\end{multline}
and similarly
\begin{multline}\label{eqn:s1_vel_spin}
A_s
\begin{bmatrix}
u_{p,1,r} \\
u_{p,1,z} \\
\Omega_{p,1,\theta}
\end{bmatrix} 
=
\begin{bmatrix}
-\frac{4\pi}{3}\Theta_{0}^{2}(\hat{R}+\hat{r}_{p}) \\[2pt] 0 \\[2pt] \frac{8\pi}{15}\Theta_{0}\Omega_{p,0,r}
\end{bmatrix}
-\begin{bmatrix}
\vec{i}'\cdot\int_{\hat{\mathcal{P}}'}\hat{\bar{\vec{u}}}\cdot\nabla\hat{\bar{\vec{u}}}\,dV \\[2pt]
\vec{k}\cdot\int_{\hat{\mathcal{P}}'}\hat{\bar{\vec{u}}}\cdot\nabla\hat{\bar{\vec{u}}}\,dV \\[2pt]
\vec{j}'\cdot\int_{\hat{\mathcal{P}}'}(\hat{\vec{x}}'-\hat{\vec{x}}_{p}^{\prime})\cross(\hat{\bar{\vec{u}}}\cdot\nabla\hat{\bar{\vec{u}}})\,dV
\end{bmatrix} \\
+\begin{bmatrix}
\int_{\hat{\mathcal{F}}'}\boldsymbol{\mathcal{V}}(\vec{i}')\cdot\mathbf{I}_{0} \,dV \\[2pt]
\int_{\hat{\mathcal{F}}'}\boldsymbol{\mathcal{V}}(\vec{k})\cdot\mathbf{I}_{0} \,dV \\[2pt]
\int_{\hat{\mathcal{F}}'}\boldsymbol{\mathcal{V}}(\vec{j}'\cross(\hat{\vec{x}}'-\hat{\vec{x}}_{p}^{\prime}))\cdot\mathbf{I}_{0} \,dV 
\end{bmatrix} \,.
\end{multline}
Here $\hat{R}=R/a$ and $\Theta_0=(\hat{R}+\hat{r}_p)u_{p,0,\theta}$ (such that $\Theta_0\vec{k}=\bs{\Theta}_{0}$).
In previous studies the corrections $u_{p,1,\theta},\bs{\Omega}_{p,1}$ and off-diagonal elements of $A_a,A_s$ were ignored.
The net contribution from the first two vectors on the right side of each of \eqref{eqn:a1_vel_spin} and \eqref{eqn:s1_vel_spin} is typically small and can be ignored, however they have been included here for completeness.

\subsection{The completed model}

Using the result of the four equations \eqref{eqn:leading_vel_spin}, \eqref{eqn:cs_vel_spin}, \eqref{eqn:a1_vel_spin} and \eqref{eqn:s1_vel_spin} our complete migration model is:
\begin{align}\label{eqn:first_order_model}
\frac{d \hat{r}_p}{d\hat{t}} &= u_{p,0,r}+\Rey_{p}u_{p,1,r} \,,& \hat{\Omega}_{p,r} &=  \Omega_{p,r,0}+\Rey_{p}\Omega_{p,r,1} \,,\\
\frac{d \hat{z}_p}{d\hat{t}} &= u_{p,0,z}+\Rey_{p}u_{p,1,z} \,,& \hat{\Omega}_{p,z} &=  \Omega_{p,z,0}+\Rey_{p}\Omega_{p,z,1}  \,,\\
\frac{d \theta_p}{d\hat{t}} &= \frac{u_{p,0,\theta}+\Rey_{p}u_{p,1,\theta}}{\hat{R}+\hat{r}_{p}} \,,& \hat{\Omega}_{p,\theta} &=  \Omega_{p,\theta,0}+\Rey_{p}\Omega_{p,\theta,1} \,.
\end{align}
Appendix \ref{sec:symmetry_reduction} describes further simplifications that can be made to \eqref{eqn:a1_vel_spin} and \eqref{eqn:s1_vel_spin} owing to symmetries about the plane $\theta'=0$.
These result in the alternative equations \eqref{eqn:a1_vel_spin_exp} and \eqref{eqn:s1_vel_spin_exp} which provide additional insight into the interplay between axial and secondary flow components and their influence on the first order terms $\mathbf{u}_{p,1}$ and $\bs{\Omega}_{p,1}$.

Given the spherical symmetry of the particle we don't need to calculate the specific angle of rotation which has occurred about its centre. 
However, the rate of rotation $\bs{\Omega}_p$ is needed as it influences the disturbance of the surrounding fluid and the $\bs{\Omega}_{p,0}$ components are needed in order to calculate the $\vec{u}_{p,1}$ components. 
Observe that the $\bs{\Omega}_{p,1}$ components are weakly coupled to $\vec{u}_{p,1}$ through the typically small, but non-zero, off-diagonal components of the matrices $A_a,A_s$.

This first order ODE model \eqref{eqn:first_order_model} assumes that the particle motion is such that the net hydrodynamic force and torque on the particle is zero at all times.
In other words, the particle instantly attains terminal velocity with respect to its current position as it moves about within the cross-section.
Obviously this fails to capture effects due to the acceleration of the particle and the immediately surrounding fluid.
However, it is not an unreasonable assumption in regimes where the particle slowly migrates across streamlines towards a stable equilibrium.
Moreover, we have found this to be a reasonably effective model in previous work so think it is worth exploring in the context of moderate Dean numbers.
The development of an efficient second order model which captures acceleration effects is the subject of ongoing work and is beyond the scope of this study.

\section{Results}\label{sec:results}

For rectangular ducts having aspect ratios $2$ and $4$ it was noted in \cite{HardingStokesBertozzi2019} that there exists a pair of stable equilibria (with vertical symmetry about $z=0$) for all of the values $\alpha\in\{1/20,2/20,3/20,4/20\}$ over the range of $\epsilon^{-1}\in[40,1280]$ that was considered.
Herein, as we increase the Dean flow parameter $K$, this continues to be the case.
Thus, it will be useful to restrict our study to the horizontal location of the stable equilibria pair and consider how this changes with respect to different parameters, especially $K$.
In this study we consider two larger particle sizes $\alpha=5/20,6/20$ in addition to those of the previous study. 
We first illustrate how the flow rate modifies the horizontal focusing location for two fixed duct bend radii corresponding to $\epsilon^{-1}=80,160$.
We then examine how long (in terms of both time and distance) it takes particles to focus and any influence $K$ has on this.
Lastly, we examine whether the parameter $\kappa$ continues to characterise the horizontal focusing location of particles for increasing Dean number.

The computations required to compute the coefficient fields described in Section~\ref{sec:qv0_expansion} were conducted using the open source finite element software FEniCS~\citep{AlnaesEtal2015,LoggEtal2012}.
The background flow was estimated over the appropriate two-dimensional cross-section by solving the equations described in \citet{HardingANZIAMJ2019}. 
Third-order Lagrange elements were used for this calculation over a triangular mesh having roughly 500,000 elements so that the relative error is expected to be much less than $10^{-6}$.
A finite difference code was used for validation of the background flow approximation and is available at \citet{HardingGitHubCDFC}.
For the three-dimensional computations of the disturbance flow, Taylor--Hood elements were used with tetrahedral meshes which were refined around the particle and typically had from 500,000 to 1,000,000 cells over the duct section.
A convergence analysis of the code was conducted in \citet{Harding2019} from which we are confident that the relative error in the computed coefficients is typically less than $10^{-3}$.
The required coefficient fields were estimates via 1000 to 2000 samples within the cross-section (depending on aspect ratio and particle size) and bivariate cubic interpolation was used for the purpose of sampling these fields within standard ODE solvers.
A Python class for accessing the computed coefficients in the case of small Dean number is available at \citet{HardingGitHubILFC} and will be updated with moderate Dean number data in the near future.

\subsection{Influence of Dean number on lateral focusing and size based separation.}\label{sec:increasing_dean_number}

\begin{figure}
\centering
\begin{subfigure}[b]{0.42\textwidth}
\centering
\includegraphics{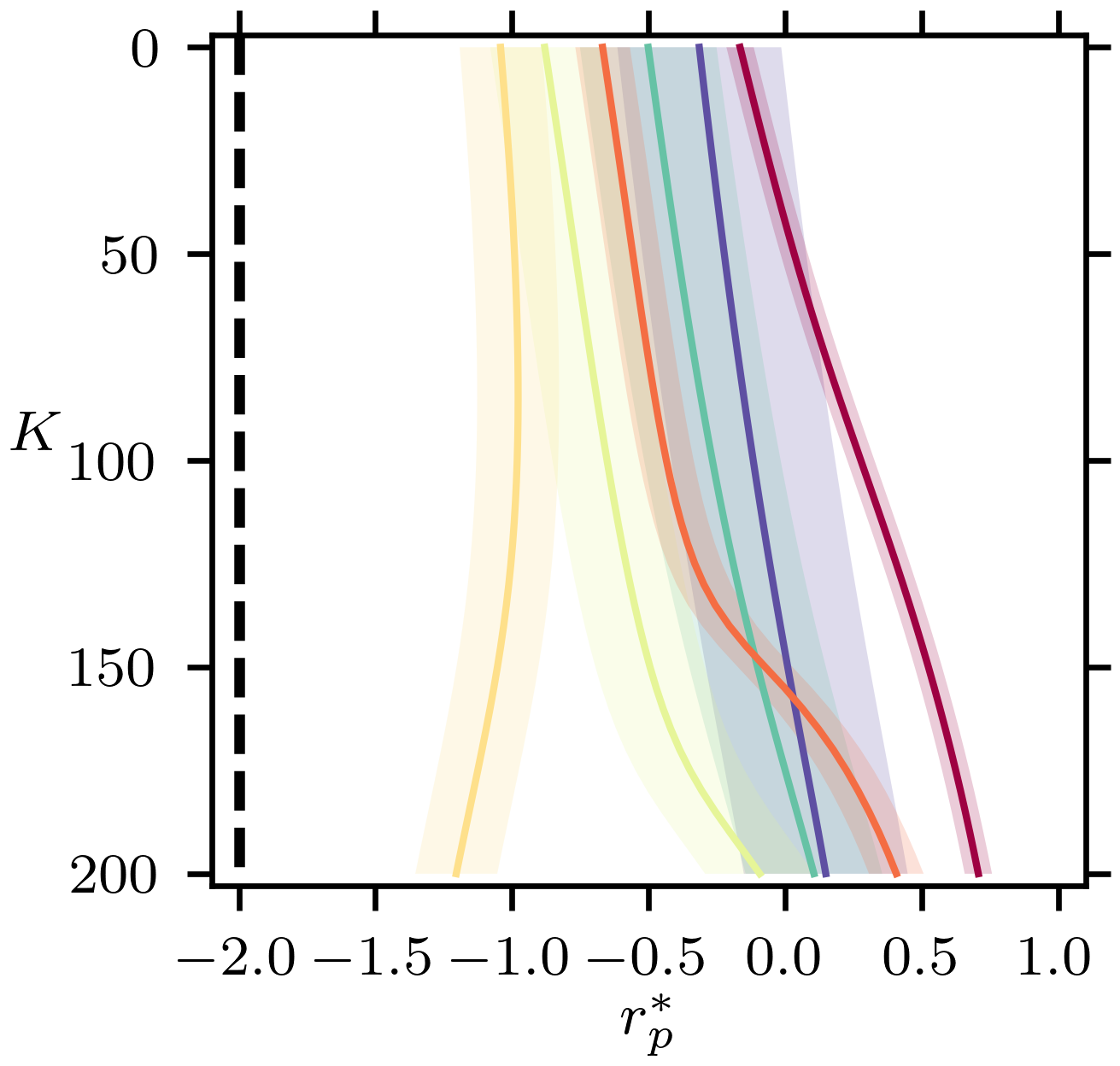}
\caption{$\epsilon^{-1}=80$}\label{fig:rect2x1_focusing_vs_Dn_a}
\end{subfigure}
\begin{subfigure}[b]{0.56\textwidth}
\centering
\includegraphics{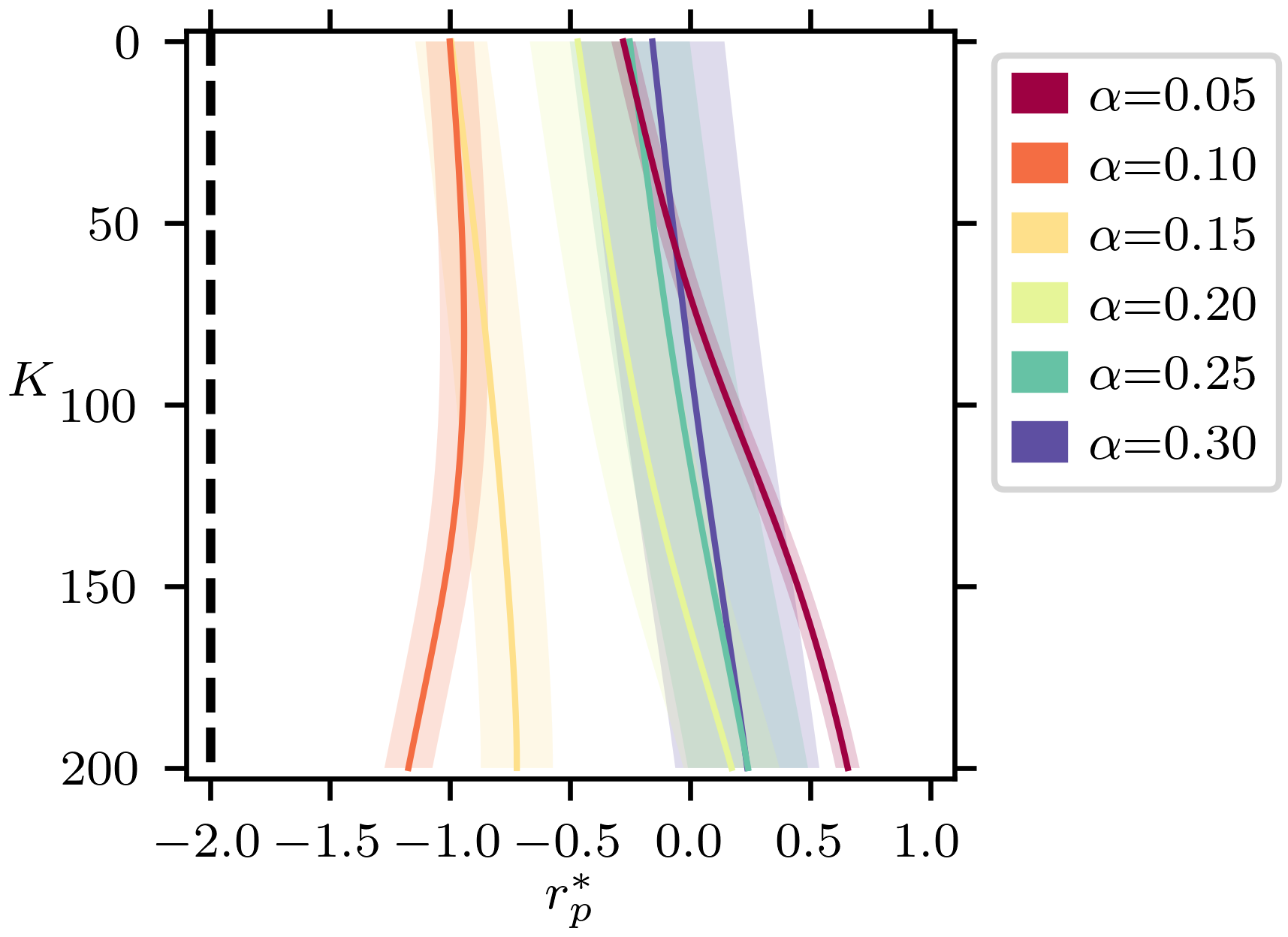}
\caption{$\epsilon^{-1}=160$}\label{fig:rect2x1_focusing_vs_Dn_b}
\end{subfigure}
\caption{Lateral particle focusing as a function of the Dean number $K$ for a cross-section with aspect ratio $2$ and bend radius (a) $\epsilon^{-1}=80$ and (b) $\epsilon^{-1}=160$. 
The horizontal focusing location, $r_{p}^{\ast}$, is non-dimensionalised with respect to $\ell/2$.
Six particles are shown with radius indicated by the shading around the solid line denoting the location of the stable equilibria.
Note the horizontal axis has been restricted to $[-2,1]$ (from $[-2,2]$).
}\label{fig:rect2x1_focusing_vs_Dn}
\end{figure}

\begin{figure}
\centering
\begin{subfigure}[b]{0.42\textwidth}
\centering
\includegraphics{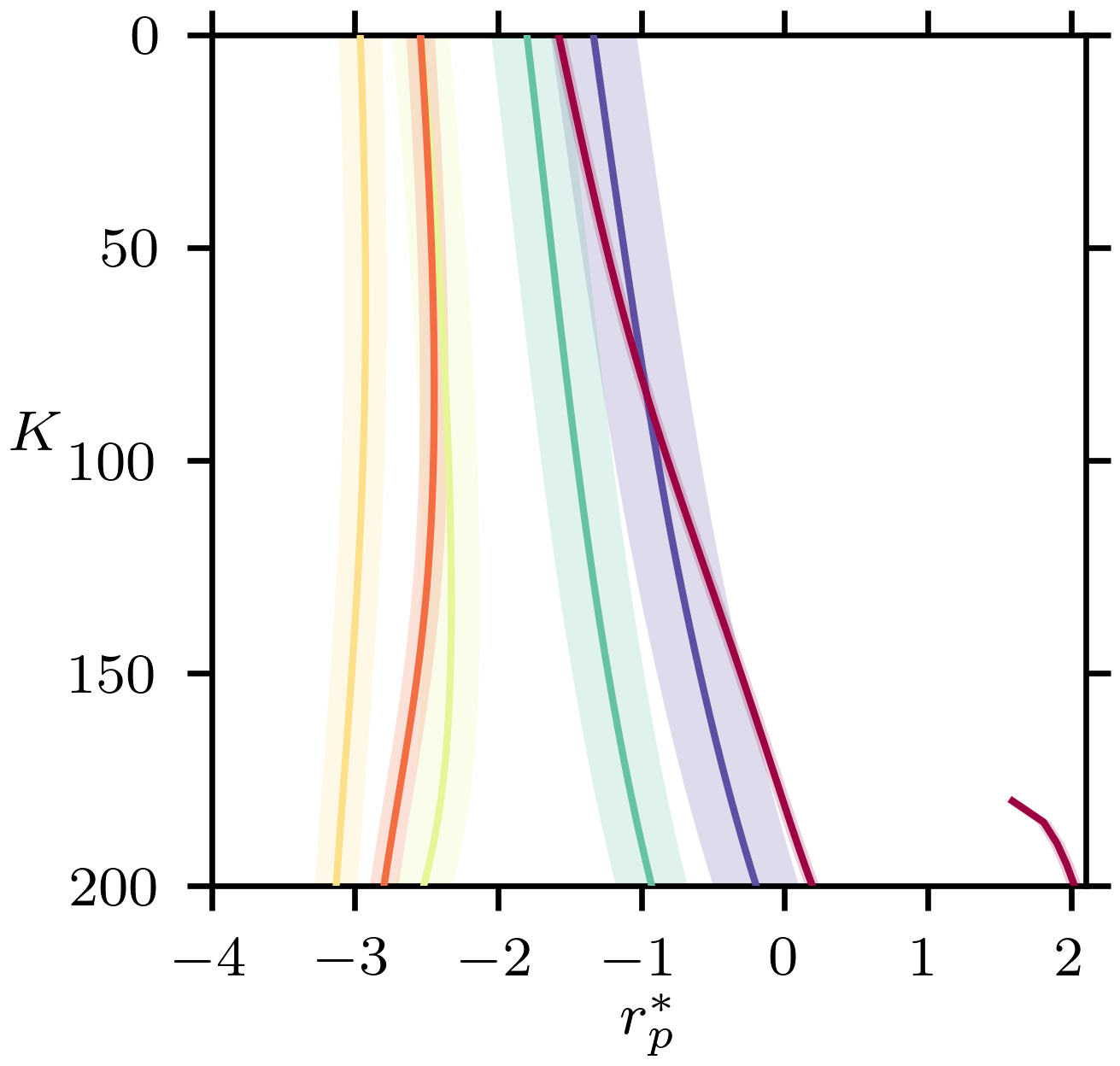}
\caption{$\epsilon^{-1}=80$}\label{fig:rect4x1_focusing_vs_Dn_a}
\end{subfigure}
\begin{subfigure}[b]{0.56\textwidth}
\centering
\includegraphics{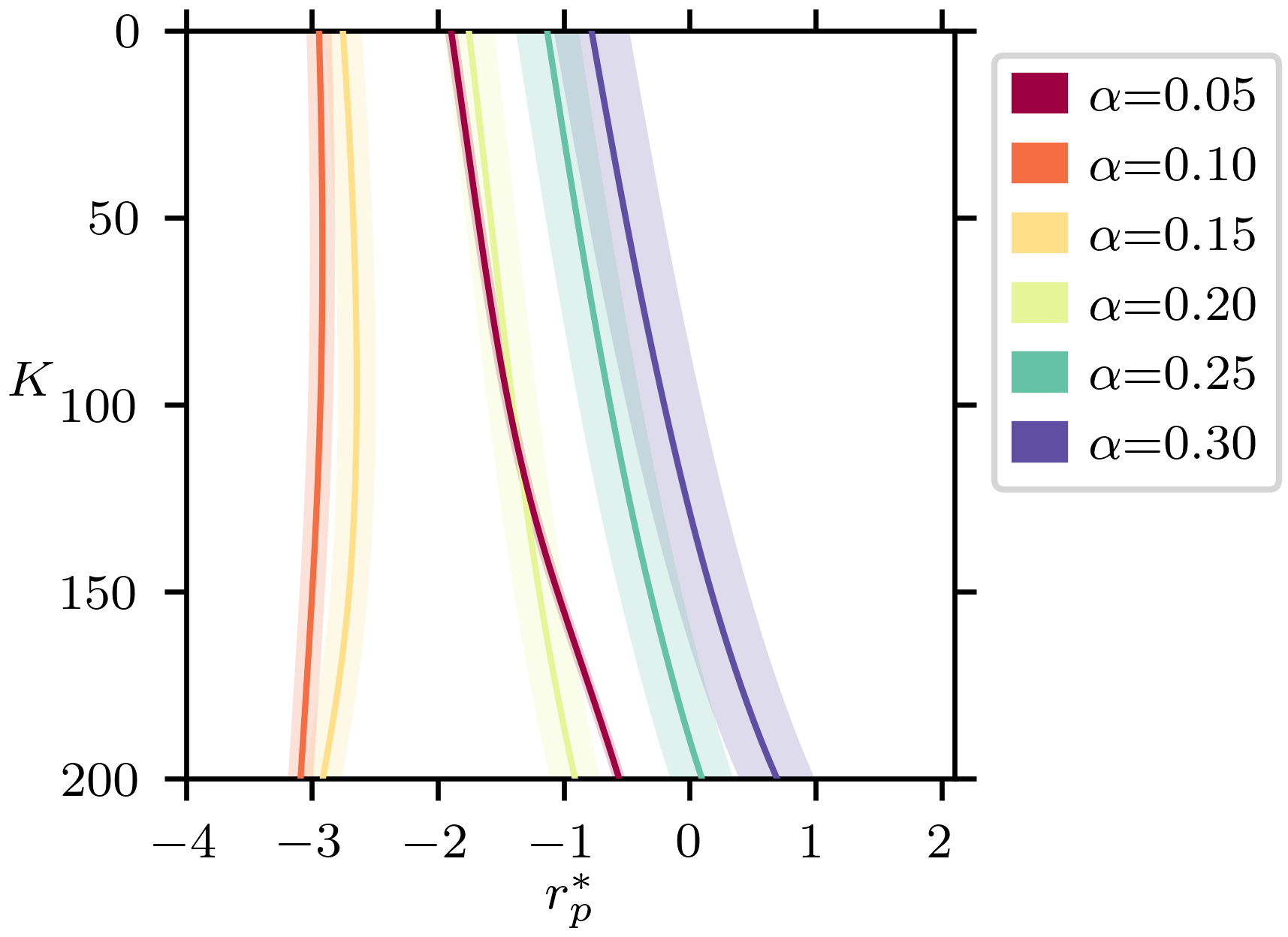}
\caption{$\epsilon^{-1}=160$}\label{fig:rect4x1_focusing_vs_Dn_b}
\end{subfigure}
\caption{Lateral particle focusing as a function of the Dean number $K$ for a cross-section with aspect ratio $4$ and bend radius (a) $\epsilon^{-1}=80$ and (b) $\epsilon^{-1}=160$.
The horizontal focusing location, $r_{p}^{\ast}$, is non-dimensionalised with respect to $\ell/2$.
Six particles are shown with radius indicated by the shading around the solid line denoting the location of the stable equilibria (for $\alpha=0.05$ this is barely perceptible). 
Note the horizontal axis has been restricted to $[-4,2]$ (from $[-4,4]$).
}\label{fig:rect4x1_focusing_vs_Dn}
\end{figure}

We investigate how the Dean number $K$ influences the focusing of particles.
Let $r_p^\ast$ denote the lateral location of a stable equilibria pair.
Figures~\ref{fig:rect2x1_focusing_vs_Dn} and \ref{fig:rect4x1_focusing_vs_Dn} illustrate how $r_p^{\ast}$ changes as $K$ increases for curved ducts with rectangular cross-sections having aspect ratio $2$ and $4$, respectively.

Considering first a duct with aspect ratio $2$ and $\epsilon^{-1}=80$.
Figure~\ref{fig:rect2x1_focusing_vs_Dn_a} illustrates a small degree of separation of the different particle sizes when $K=0$ (interpreted as the behaviour in the limit $K\to0$ as a result of a decaying flow rate).
As $K$ increases, the particles with $\alpha=0.05,0.15$ become increasingly separated from the others. 
The stable equilibria for $\alpha=0.15$ initially has $r_p^\ast\approx-1$, and initially shifts slightly towards the centre as $K$ increases before swinging back towards the left/inside wall.
The stable equilibria for $\alpha=0.05$ begins with $r\approx-0.16$ and steadily shifts towards the right (outside wall), up to $r\approx 0.67$.
In contrast, the remaining particle sizes begin with centre between the other two and shift towards slightly right of centre without spreading significantly. 
Interestingly, the $\alpha=0.10$ particle experiences a relatively rapid shift toward the outside wall around $K=150$, enough to give it some small separation from the $\alpha=0.20$ particle.

Figure~\ref{fig:rect2x1_focusing_vs_Dn_b} illustrates the case with the larger bend radius $\epsilon^{-1}=160$.
The $\alpha=0.10,0.15$ particles begin separated from the others when $K=0$.
As $K$ increases, so does the separation between the two groups.
Moreover, the $\alpha=0.10$ particle becomes slightly separated from that with $\alpha=0.15$ for $K>150$ as it shifts towards the inside wall,
Similar to the $\epsilon^{-1}=80$ case,  the $\alpha=0.05$ particle becomes slightly separated from the other group as it more readily migrates towards the outside wall.
From a practical perspective it is useful that the clear separation between the initial two groups of particles is maintained for increasing $K$ in this particular case.

We now examine the results for a duct with aspect ratio $4$ shown in Figure~\ref{fig:rect4x1_focusing_vs_Dn}. 
With $\epsilon^{-1}=80$, the $\alpha=0.10,0.15,0.20$ particles are closer to the inside wall than the others at $K=0$.
This separation increases for increasing $K$, largely due to $r_p^\ast$ for those $\alpha=0.05,0.25,0.30$ particles shifting towards the outside wall.
Additionally, each of the $\alpha=0.05,0.25,0.30$ particles attains a small degree of separation from each other for $K$ approaching $200$.
The $\alpha=0.15$ particle also achieves some separation from the $\alpha=0.10,0.20$ particles over the entire range of $K$, with maximum separation when $K\approx125$.
For $\alpha=0.05$, as $K$ approaches $200$ there is a bifurcation that leads to a second pair of stable equilibria in the half of the duct nearer to the outside wall.

With $\epsilon^{-1}=160$, there are three distinct groups when $K=0$, namely $\alpha=0.10,0.15$, $\alpha=0.05,0.20$ and $\alpha=0.25,0.30$ ordered from the inside wall to the outside wall.
The separation between the three groups is maintained for increasing $K$,  with the first group becomes increasingly separated from the latter two.

All together, these results suggest several significant trends.
First, clear separation between the stable equilibria of groups of different particle sizes at a low flow rate is not adversely affected as the Dean number increases.
Second, particle sizes which focus near the left wall when $K=0$ don't move significantly for increasing $K$ and, as a consequence, become increasingly separated from the other particle sizes which shift towards the outside wall.
Third, the $\alpha=0.05$ particle tends to focus near the larger particles but can achieve some separation for large enough $K$.
The $\alpha=0.15$ particle appears to be well separated from the remainder of the particles most consistently across these results.

\subsection{Trajectory illustrations}\label{sec:trajectory_plots}

We provide illustrations of particle trajectories towards stable equilibria for some selected values of $\alpha,\epsilon^{-1},K$.
These help to demonstrate the effect of increasing Dean number on the migration dynamics of particles and assist with the interpretation of the results from Section~\ref{sec:increasing_dean_number}.

\begin{figure}
\centering
\begin{subfigure}[b]{0.32\textwidth}
\centering
\includegraphics{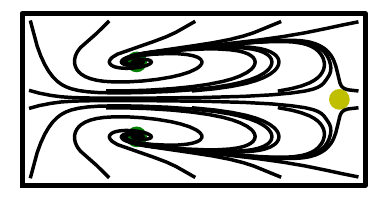}
\caption{$\alpha=0.10$, $K=1$}\label{fig:traj_2x1_R80_a}
\end{subfigure}
\begin{subfigure}[b]{0.32\textwidth}
\centering
\includegraphics{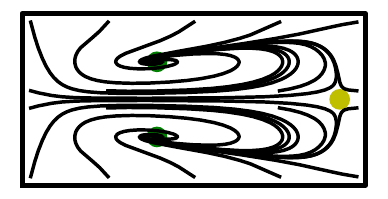}
\caption{$\alpha=0.10$, $K=100$}\label{fig:traj_2x1_R80_b}
\end{subfigure}
\begin{subfigure}[b]{0.32\textwidth}
\centering
\includegraphics{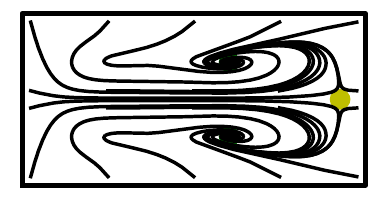}
\caption{$\alpha=0.10$, $K=200$}\label{fig:traj_2x1_R80_c}
\end{subfigure}
\\
\begin{subfigure}[b]{0.32\textwidth}
\centering
\includegraphics{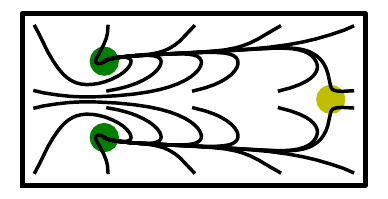}
\caption{$\alpha=0.15$, $K=1$}\label{fig:traj_2x1_R80_d}
\end{subfigure}
\begin{subfigure}[b]{0.32\textwidth}
\centering
\includegraphics{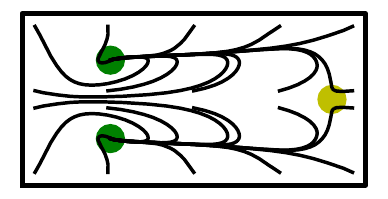}
\caption{$\alpha=0.15$, $K=100$}\label{fig:traj_2x1_R80_e}
\end{subfigure}
\begin{subfigure}[b]{0.32\textwidth}
\centering
\includegraphics{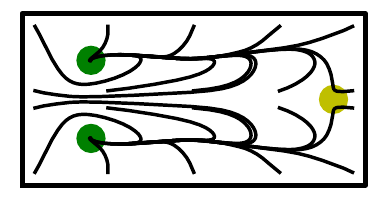}
\caption{$\alpha=0.15$, $K=200$}\label{fig:traj_2x1_R80_f}
\end{subfigure}
\caption{Trajectories of particles towards stable equilibria in a curved rectangular duct with aspect ratio $2$ and dimensionless bend radius $\epsilon^{-1}=80$.
The particle has size (a,b,c) $\alpha=0.10$ and (d,e,f) $\alpha=0.15$, and the Dean number has values (a,d) $K=1$, (b,e) $K=100$ and (c,f) $K=200$.
The left side is the inside wall of the curved duct.
Stable equilibria are green, saddle equilibria are yellow and unstable equilibria are red.
The marker size reflects the size of the particle.
}\label{fig:traj_2x1_R80}
\end{figure}

Figure~\ref{fig:traj_2x1_R80} illustrates trajectories of two distinct particle sizes at three different Dean numbers in a curved duct with aspect ratio 2 and dimensionless bend radius $\epsilon^{-1}=80$.
The stable equilibria pair for the $\alpha=0.10$ particle shifts to the right as $K$ increases (as also evident in Figure~\ref{fig:rect2x1_focusing_vs_Dn_a}) and the trajectories deform accordingly.
Conversely, the equilibria pair of the slightly larger $\alpha=0.15$ particle shifts very little over this range of $K$. 
For increasing $K$, particles starting near the inside (left) wall migrate further to the right before making their way back to a stable equilibrium along the slow manifold.
Additionally, when $K=200$ we observe a slight deformation of the slow manifold, evident as a pinch near its centre.
Observe that the horizontal location of the stable equilibria for the two particle sizes has only a small degree of separation when $K=1$ (compare \ref{fig:traj_2x1_R80_a} and \ref{fig:traj_2x1_R80_d}) but increases with increasing $K$ (compare \ref{fig:traj_2x1_R80_c} and \ref{fig:traj_2x1_R80_f}).

\begin{figure}
\centering
\begin{subfigure}[b]{0.32\textwidth}
\centering
\includegraphics{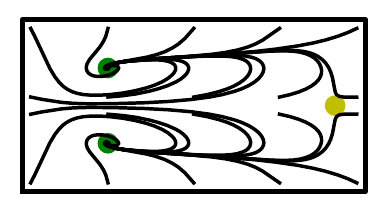}
\caption{$\alpha=0.10$, $K=1$}\label{fig:traj_2x1_R160_a}
\end{subfigure}
\begin{subfigure}[b]{0.32\textwidth}
\centering
\includegraphics{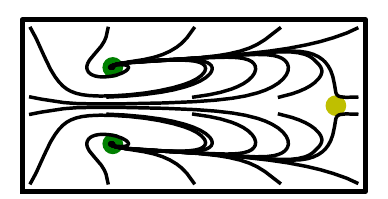}
\caption{$\alpha=0.10$, $K=100$}\label{fig:traj_2x1_R160_c}
\end{subfigure}
\begin{subfigure}[b]{0.32\textwidth}
\centering
\includegraphics{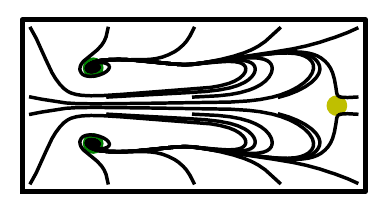}
\caption{$\alpha=0.10$, $K=200$}\label{fig:traj_2x1_R160_e}
\end{subfigure}
\\
\begin{subfigure}[b]{0.32\textwidth}
\centering
\includegraphics{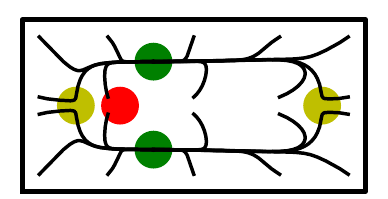}
\caption{$\alpha=0.20$, $K=1$}\label{fig:traj_2x1_R160_b}
\end{subfigure}
\begin{subfigure}[b]{0.32\textwidth}
\centering
\includegraphics{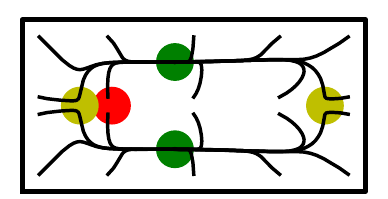}
\caption{$\alpha=0.20$, $K=100$}\label{fig:traj_2x1_R160_d}
\end{subfigure}
\begin{subfigure}[b]{0.32\textwidth}
\centering
\includegraphics{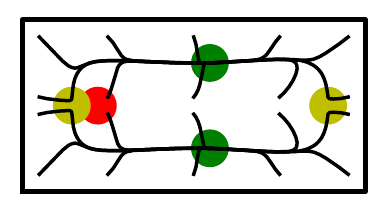}
\caption{$\alpha=0.20$, $K=200$}\label{fig:traj_2x1_R160_f}
\end{subfigure}
\caption{Trajectories of particles towards stable equilibria in a curved rectangular duct with aspect ratio $2$ and dimensionless bend radius $\epsilon^{-1}=160$.
The particles have sizes (a,b,c) $\alpha=0.10$ and (d,e,f) $\alpha=0.20$, and the Dean number has values (a,d) $K=1$, (b,e) $K=100$ and (c,f) $K=200$.
The left side is the inside wall of the curved duct.
Stable equilibria are green, saddle equilibria are yellow and unstable equilibria are red.
The marker size reflects the size of the particle.
}\label{fig:traj_2x1_R160}
\end{figure}

Figure~\ref{fig:traj_2x1_R160} similarly illustrates trajectories of two distinct particle sizes at three different Dean numbers in a curved duct with aspect ratio 2, but with the dimensionless bend radius $\epsilon^{-1}=160$.
The stable equilibria pair for the $\alpha=0.10$ particle doesn't shift much as $K$ increases (as also evident in Figure~\ref{fig:rect2x1_focusing_vs_Dn_b}), but particles starting near the inside (left) wall migrate further to the right before making their way back to a stable equilibrium along the slow manifold.
Additionally, when $K=200$ we again observe a slight pinching deformation of the slow manifold.
This is much like the behaviour when $\alpha=0.15$ and $\epsilon^{-1}=80$ from Figure~\ref{fig:traj_2x1_R80}.
For the larger $\alpha=0.20$ particle, the stable equilibria pair shifts to the right with increasing $K$ but with only subtle changes to the trajectories as a whole.
Again we observe that the horizontal separation of stable equilibria between the two particle sizes increases with increasing $K$, however in this case it is the larger particle that shifts towards the outer wall, unlike the case in Figure~\ref{fig:traj_2x1_R80}.

\begin{figure}
\centering
\begin{subfigure}[b]{0.48\textwidth}
\centering
\includegraphics{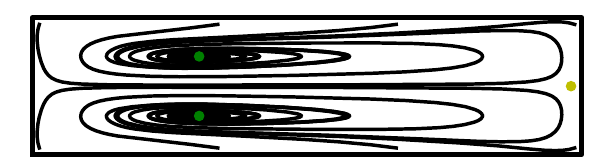}
\caption{$\alpha=0.05$, $K=1$}\label{fig:traj_4x1_R80_a}
\end{subfigure}
\begin{subfigure}[b]{0.48\textwidth}
\centering
\includegraphics{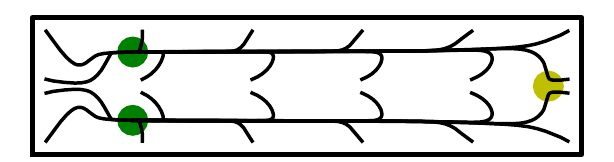}
\caption{$\alpha=0.20$, $K=1$}\label{fig:traj_4x1_R80_b}
\end{subfigure}
\\
\begin{subfigure}[b]{0.48\textwidth}
\centering
\includegraphics{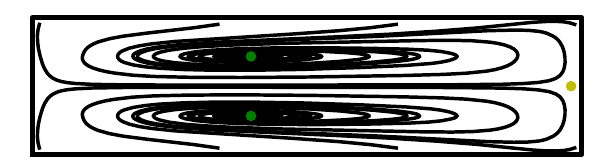}
\caption{$\alpha=0.05$, $K=100$}\label{fig:traj_4x1_R80_c}
\end{subfigure}
\begin{subfigure}[b]{0.48\textwidth}
\centering
\includegraphics{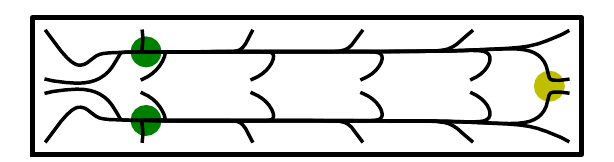}
\caption{$\alpha=0.20$, $K=100$}\label{fig:traj_4x1_R80_d}
\end{subfigure}
\\
\begin{subfigure}[b]{0.48\textwidth}
\centering
\includegraphics{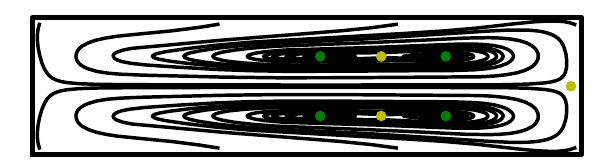}
\caption{$\alpha=0.05$, $K=200$}\label{fig:traj_4x1_R80_e}
\end{subfigure}
\begin{subfigure}[b]{0.48\textwidth}
\centering
\includegraphics{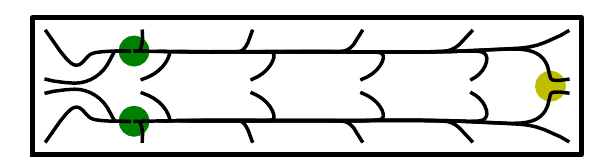}
\caption{$\alpha=0.20$, $K=200$}\label{fig:traj_4x1_R80_f}
\end{subfigure}
\caption{Trajectories of particles towards stable equilibria in a curved rectangular duct with aspect ratio $2$ and dimensionless bend radius $\epsilon^{-1}=80$.
The particle size is (a,c,e) $\alpha=0.05$ and (b,d,f) $\alpha=0.20$, and the Dean number has values (a,b) $K=1$, (c,d) $K=100$ and (e,f) $K=200$.
The left side is the inside wall of the curved duct.
Stable equilibria are green, saddle equilibria are yellow and unstable equilibria are red.
The marker size reflects the size of the particle.
}\label{fig:traj_4x1_R80}
\end{figure}

Figure~\ref{fig:traj_4x1_R80} illustrates trajectories of two distinct particle sizes at three different Dean numbers in a curved duct with aspect ratio $4$ and dimensionless bend radius $\epsilon^{-1}=80$.
The stable equilibria for the smaller $\alpha=0.05$ particle shifts to the right with increasing $K$ and the spiral trajectories become increasingly elongated.
For $K=200$ there is an additional pair of stable equilibria located nearer to the outer wall (as also seen in Figure~\ref{fig:rect4x1_focusing_vs_Dn_a}) and, although not easily seen in the figure, this pair captures more of the particles than the stable pair located closer to the centre.
Given that our secondary flow velocity approximation has an accuracy of $O(10)\%$ when $K=200$ it is unclear if a two pairs of stable equilibria occurs in practice (and is difficult to verify since fluorescent streak images from experiments typically show that small particles have not converged \citep{RafeieEtal2019}).
For the larger $\alpha=0.20$ particle there is almost no perceptible change in the migration dynamics over this range of $K$ values.

\begin{figure}
\centering
\begin{subfigure}[b]{0.48\textwidth}
\centering
\includegraphics{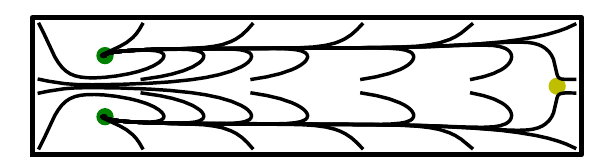}
\caption{$\alpha=0.10$, $K=1$}\label{fig:traj_4x1_R160_a}
\end{subfigure}
\begin{subfigure}[b]{0.48\textwidth}
\centering
\includegraphics{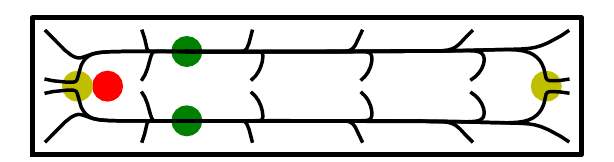}
\caption{$\alpha=0.20$, $K=1$}\label{fig:traj_4x1_R160_b}
\end{subfigure}
\\
\begin{subfigure}[b]{0.48\textwidth}
\centering
\includegraphics{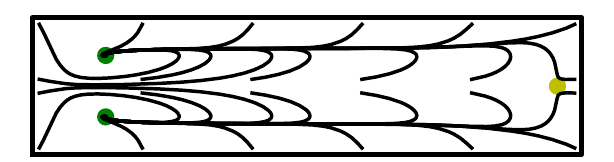}
\caption{$\alpha=0.10$, $K=100$}\label{fig:traj_4x1_R160_c}
\end{subfigure}
\begin{subfigure}[b]{0.48\textwidth}
\centering
\includegraphics{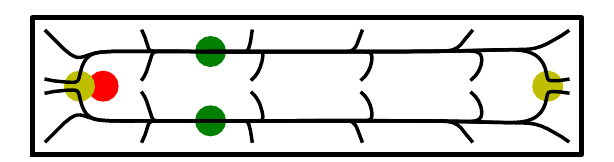}
\caption{$\alpha=0.20$, $K=100$}\label{fig:traj_4x1_R160_d}
\end{subfigure}
\\
\begin{subfigure}[b]{0.48\textwidth}
\centering
\includegraphics{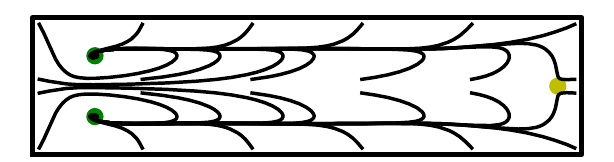}
\caption{$\alpha=0.10$, $K=200$}\label{fig:traj_4x1_R160_e}
\end{subfigure}
\begin{subfigure}[b]{0.48\textwidth}
\centering
\includegraphics{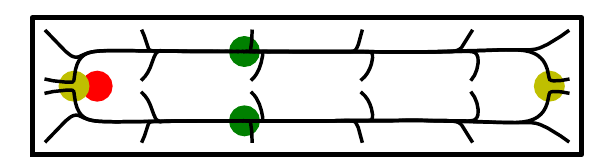}
\caption{$\alpha=0.20$, $K=200$}\label{fig:traj_4x1_R160_f}
\end{subfigure}
\caption{Trajectories of particles towards stable equilibria in a curved rectangular duct with aspect ratio $2$ and dimensionless bend radius $\epsilon^{-1}=160$.
The particle size is (a,c,e) $\alpha=0.10$ and (b,d,f)$\alpha=0.20$ , and the Dean number has values (a,b) $K=1$, (c,d) $K=100$ and (e,f) $K=200$.
The left side is the inside wall of the curved duct.
Stable equilibria are green, saddle equilibria are yellow and unstable equilibria are red.
The marker size reflects the size of the particle.
}\label{fig:traj_4x1_R160}
\end{figure}

Figure~\ref{fig:traj_4x1_R160} illustrates trajectories of two distinct particle sizes at three different Dean numbers in a curved duct with aspect ratio 4, but with dimensionless bend radius $\epsilon^{-1}=160$.
We observe the dynamics of the smaller $\alpha=0.10$ particle changes very little over this range of $K$. 
The stable equilibria pair for the larger $\alpha=0.20$ particle shifts towards the centre along the slow manifold for increasing $K$ but with almost no perceptible change in the dynamics of trajectories otherwise.
Note that the horizontal separation between the larger and smaller particle again increases with $K$.

\subsection{Focusing time and distance}\label{sec:focus_time}

Here we investigate the time and axial distance required for particles to focus and whether the Dean number has a significant effect on this.
The migration time $t_{m}(r,z)$ of an individual particle starting with centre at $(r_{p}(0),z_{p}(0))=(r,z)\in\mathcal{C}$ (such that the particle does not intersect the duct walls), is taken to be the earliest time after which the particle remains within $0.01\ell$ of a stable equilibrium, specifically
\begin{equation}\label{eqn:focus_time}
t_{m}(r,z) := \min\left\{t\geq0 \,\middle|
\begin{array}{l}(r_p(0),z_p(0))=(r,z)\text{ and} \\ 
\|(r_p(\tau),z_p(\tau))-(r_p^\ast,z_p^\ast)\|\leq\ell/100 \text{ for all }\tau\geq t\,\end{array}\right\}\,.
\end{equation}
Here $(r_p^\ast,z_p^\ast)$ is taken to be the specific equilibrium which $(r_p(\tau),z_p(\tau))$ converges towards as $\tau\to\infty$.
At the focusing time $t_m=t_m(r,z)$, we also record the total angle around the curved duct travelled by the particle, that is $\theta_p(t_m)$, and from this we calculate a projected axial distance $d_m=R\theta_p(t_m)$.

Observe that \eqref{eqn:focus_time} is a slight modification of the definition in \citet{HaEtal22}. 
We make the distinction of ensuring the particle remains within $0.01\ell$ of a stable equilibrium for all $\tau\geq t$ because a small particle, for which the secondary flow drag is dominant, may migrate within $0.01\ell$ of $(r_p^\ast,z_p^\ast)$ and then away again as it follows an elliptical shaped spiral towards it.
This wasn't an issue in the prior square duct study because the background flow vortices are not as elongated as they are within the rectangular ducts considered here.

For a given triple $\alpha,\epsilon,K$ we determine the migration time as the time at which $90\%$ of particles are determined to be focused.
This is estimated by calculating $t_m$ for a large number of individual particles with starting locations defining a grid of equidistant points in $\mathcal{C}$.
From these times we then select the 90th percentile.
Similarly, the 90th percentile is selected from the migration distances $d_m$.
Here we present results using the dimensionless migration time $\tilde{t}_m$ and distance $\tilde{d}_m$ which are related to their physical quantities via $t_m=\tilde{t}_m(\ell/U_m)(\ell/a)\Re_{p}^{-1}$ and $d_m=\tilde{d}_m \ell(\ell/a)\Re_{p}^{-1}$, respectively.

\begin{figure}
\centering
\begin{subfigure}[b]{0.42\textwidth}
\centering
\includegraphics{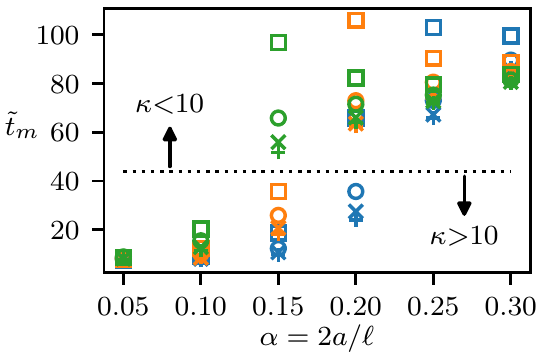}
\caption{Focusing time vs particle size}\label{fig:focusing_time_a}
\end{subfigure}
\begin{subfigure}[b]{0.57\textwidth}
\centering
\includegraphics{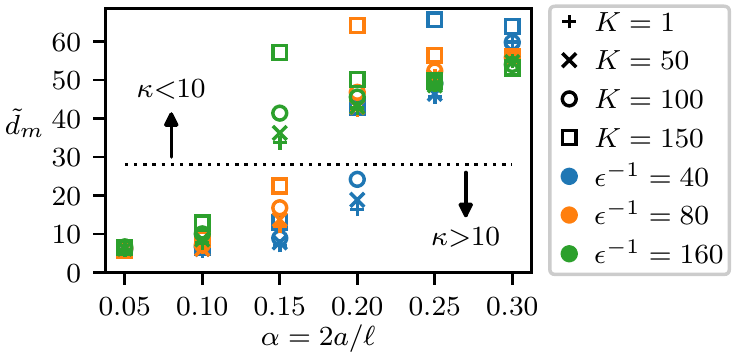}
\caption{Focusing distance vs particle size}\label{fig:focusing_dist_a}
\end{subfigure}
\\
\begin{subfigure}[b]{0.42\textwidth}
\centering
\includegraphics{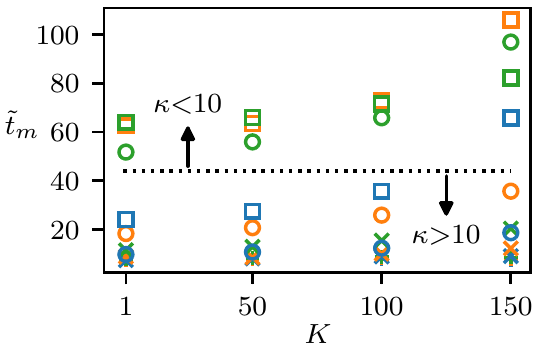}
\caption{Focusing time vs Dean number}\label{fig:focusing_time_b}
\end{subfigure} 
\begin{subfigure}[b]{0.57\textwidth}
\centering
\includegraphics{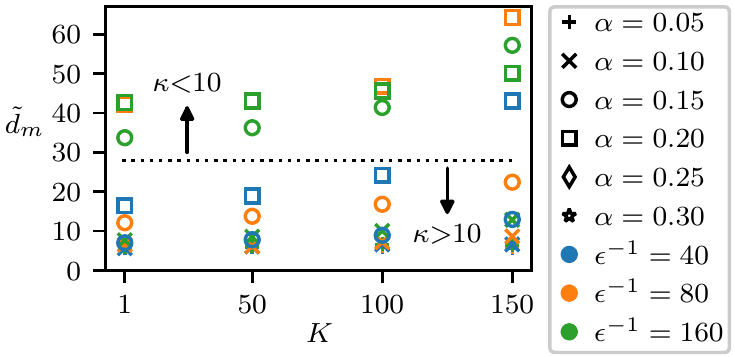}
\caption{Focusing distance vs Dean number}\label{fig:focusing_dist_b}
\end{subfigure}
\caption{
Focusing time $\tilde{t}_m$ (a,c) and distance $\tilde{d}_m$ (b,d) versus particle size $\alpha$ (a,b) and Dean number $K$ (c,d). 
The cross-section has aspect ratio 2.
In (a,b) the colours and markers differentiate between $\epsilon^{-1}$ and $K$ values respectively.
In (c,d) the colours and markers differentiate between $\epsilon^{-1}$ and $\alpha$ values respectively.
Below the black dotted lines all data points satisfy $\kappa>10$, while above all but one satisfies $\kappa<10$.
}\label{fig:focusing_time}
\end{figure}

The results of these focusing time and distance measurements for a cross-section with aspect ratio $2$ are depicted in Figure~\ref{fig:focusing_time}.
The focusing times and distances are qualitatively similar, so we will focus our discussion on the focusing times.

Two distinct regimes are identified in the figures either side of the black dotted line: data for which $\kappa>10$ is below; while data with $\kappa<10$ is above; excepting one data point with $\alpha=0.20$, $\epsilon^{-1}=40$ and $K=150$.
The boundary $\kappa=10$ roughly distinguishes between when the inertial lift is dominant ($\kappa\ll10$) and the secondary flow drag is dominant ($\kappa\gg10$).
We observe that the $\kappa>10$ regime generally appears to focus much quicker.
This is most likely explained by the absence of a slow manifold in the cross-sectional plane when the secondary flow drag is dominant.
The triple $\alpha,\epsilon^{-1},K=0.2,40,150$, an outlier for which $\kappa>10$ but is located above the dotted line, appears to migrate slower than anticipated as it is near a bifurcation (with respect to $K$) which gives rise to two stable pairs within the cross-section. 
We note that samples for $K=200$ were also calculated but ultimately excluded from these figures due to many similar outliers.

The results also demonstrate that the focusing time and distance typically appears to increase slightly with increasing $\alpha$. 
This can be explained by the (dimensionless) drag coefficient increasing with $\alpha$ (as predicted by Faxen's law), and also the inertial lift force decreasing slightly in magnitude as $\alpha$ increases. 
The combination of these two effects contributes to a decreasing inertial migration velocity with respect to $\alpha$, thereby leading to larger focusing times.

Additionally, we observe that increasing $K$ typically results in an increasing spread of focusing times.
Moreover, in the regime $\kappa<10$, the focusing times generally appear to increase with increasing $K$.
A possible explanation is that for sufficiently large $K$ we sometimes observe a bifurcation resulting in multiple stable focusing pairs.
As $K$ approaches this bifurcation points there is a location on the slow manifold where the migration velocity tends towards zero thereby greatly increasing the focusing time of particles that must pass through this region.

\begin{figure}
\centering
\begin{subfigure}[b]{0.42\textwidth}
\centering
\includegraphics{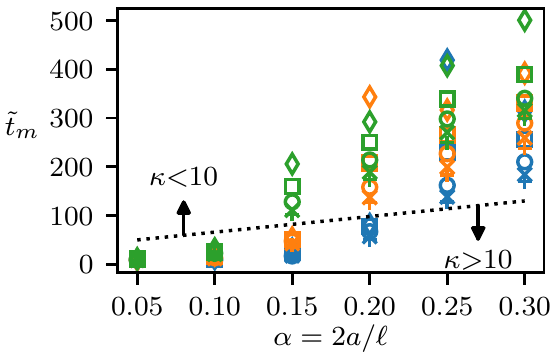}
\caption{Focusing time vs particle size}\label{fig:focusing_time2_a}
\end{subfigure}
\begin{subfigure}[b]{0.57\textwidth}
\centering
\includegraphics{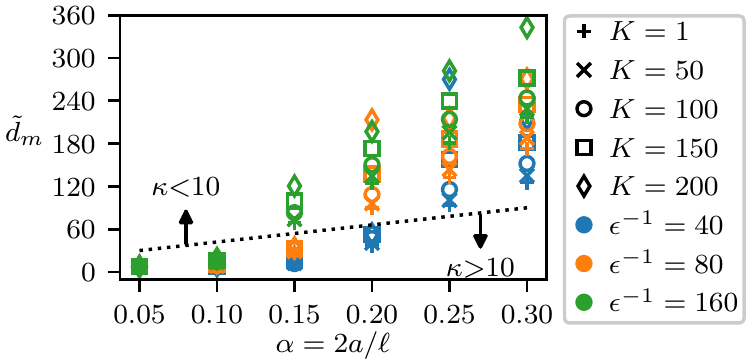}
\caption{Focusing distance vs particle size}\label{fig:focusing_dist2_a}
\end{subfigure}
\\
\begin{subfigure}[b]{0.42\textwidth}
\centering
\includegraphics{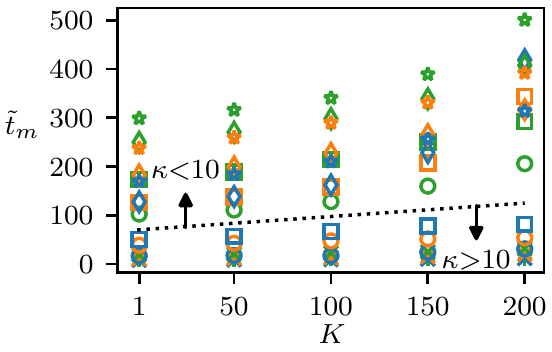}
\caption{Focusing time vs Dean number}\label{fig:focusing_time2_b}
\end{subfigure}
\begin{subfigure}[b]{0.57\textwidth}
\centering
\includegraphics{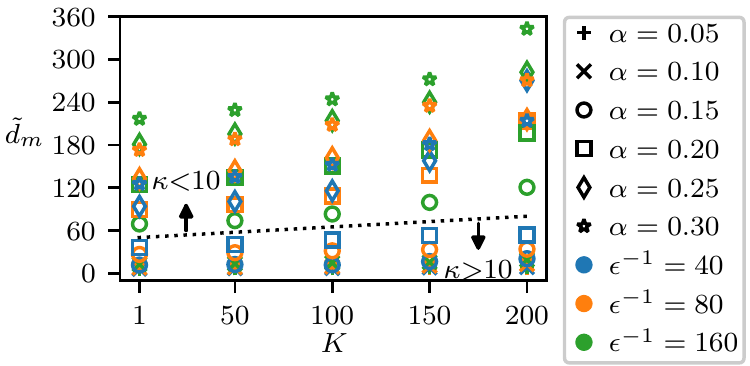}
\caption{Focusing distance vs Dean number}\label{fig:focusing_dist2_b}
\end{subfigure}
\caption{
Focusing time $\tilde{t}_m$ (a,c) and distance $\tilde{d}_m$ (b,d) versus particle size $\alpha$ (a,b) and Dean number $K$ (c,d). 
The cross-section has aspect ratio 4.
Above the black dotted line all data points satisfy $\kappa\leq10$, while below they satisfy $\kappa>10$.
In (a,b) the colours and markers differentiate between $\epsilon^{-1}$ and $K$ values respectively.
In (c,d) the colours and markers differentiate between $\epsilon^{-1}$ and $\alpha$ values respectively.
}\label{fig:focusing_time2}
\end{figure}

Figure~\ref{fig:focusing_time2} illustrates the focusing times and distances within a cross-section having aspect ratio $4$.
The observations are generally the same as the aspect ratio $2$ case.
We again use a black dotted line to differentiate data points with $\kappa<10$ above and $\kappa>10$ below (with no outliers in this instance).  
Although the gap between the two subsets of the data is smaller in this case there is a small degree of separation.
Observe that the focusing times for data with $\kappa<10$ are much larger than those seen in the duct with aspect ratio 2.
This occurs because the particles have further to migrate due to the increased duct width and, moreover, much of this extra distance is covered along the slow manifold.
Analogous observations are made with respect to the focusing distance.

\subsection{Comparison with an experimental study}

\citet{RafeieEtal2019} studied particle focusing in a spiral duct with rectangular cross-section having aspect ratio $4$ (designated design R3).
For backward flow (going from the innermost to the outermost turn) $\epsilon^{-1}\approx 180$ near the outlet, while for forward flow (from the outermost to the innermost turn) $\epsilon^{-1}\approx 74$ near the outlet. 
These two values are reasonably close to the two values $\epsilon^{-1}=160,80$ presented in our study.
They experimented with three particle sizes $\alpha\approx0.04,0.067,0.1$ and used flow rates from $0.5$ to $9.0$ mL$/$min in steps of $0.5$ mL$/$min (corresponding to Dean numbers from $K\approx 2$ up to $840$ at the innermost turn).
Their results suggest the smallest particle remains unfocused for both flow directions, and appears to be distributed more uniformly for larger flow rates. 
On the other hand, the larger two particle sizes start focused near the inside wall at low flow rates and shift significantly towards the outside wall as the flow rate increases.

Bearing in mind that $K=200$ only covers up to a flow rate of approximately $5.0$ mL$/$min for this specific device (with $K$ measured on the innermost turn), we observe that the shifting of the two larger particle sizes towards the outside wall is qualitatively similar behaviour to our predictions for a $\alpha=0.05$ particle.
Curiously, our $\alpha=0.10$ results suggest this particle stays near the inside wall, something that is not observed in the experiments for this particle size.
In the case of the smallest particle in the experiment ($\alpha\approx0.04$), our model predicts that the duct is much too short, by roughly a factor of $10$, to achieve full focusing (see section~\ref{sec:focus_time}), and this is supported by the wide spread of streaks.
Moreover, the amount of spread in the experimental results for $\alpha\approx0.067$ suggests the duct is too short for complete focusing of this particle size as well. 
This too is supported by our model which predicts the duct needs to be roughly $3$ times longer.

\subsection{Approximate collapse of horizontal focusing location}

In \citet{HardingStokesBertozzi2019} it was demonstrated that at low flow rates the horizontal location of stable equilibria pairs approximately collapses when plotted against $\kappa$.
Here we investigate whether this remains the case when the Dean number increases.

\begin{figure}
\centering
\begin{subfigure}[b]{0.42\textwidth}
\centering
\includegraphics{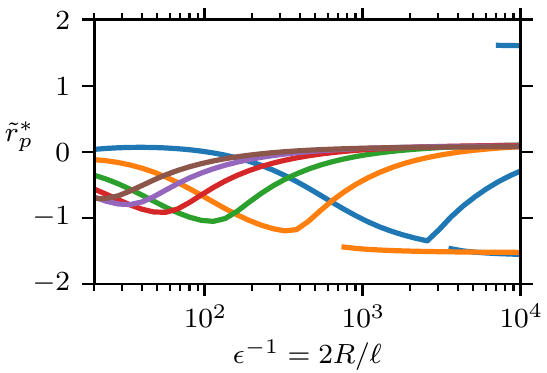}
\caption{$\tilde{r}_p^\ast$ vs $\epsilon^{-1}$, $K=50$}
\end{subfigure}
\begin{subfigure}[b]{0.56\textwidth}
\centering
\includegraphics{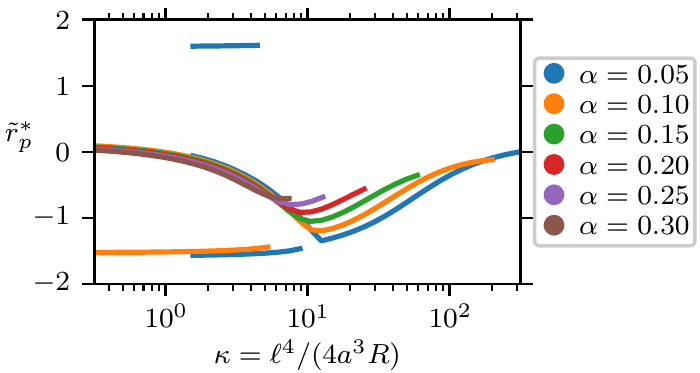}
\caption{$\tilde{r}_p^\ast$ vs $\kappa$, $K=50$}
\end{subfigure}
\\
\begin{subfigure}[b]{0.42\textwidth}
\centering
\includegraphics{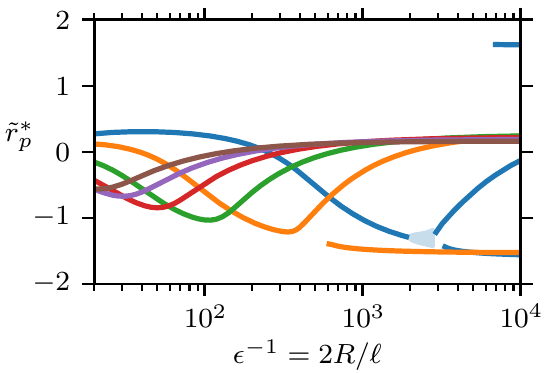}
\caption{$\tilde{r}_p^\ast$ vs $\epsilon^{-1}$, $K=100$}
\end{subfigure}
\begin{subfigure}[b]{0.56\textwidth}
\centering
\includegraphics{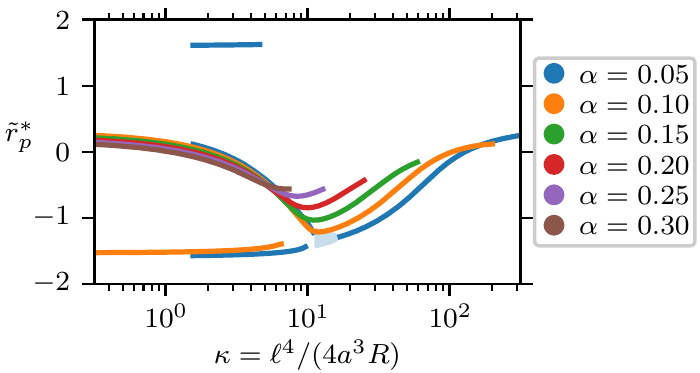}
\caption{$\tilde{r}_p^\ast$ vs $\kappa$, $K=100$}
\end{subfigure}
\\
\begin{subfigure}[b]{0.42\textwidth}
\centering
\includegraphics{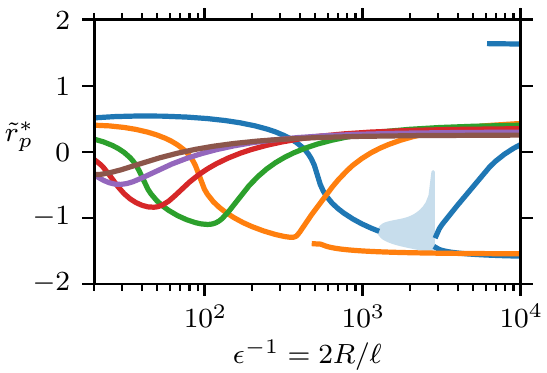}
\caption{$\tilde{r}_p^\ast$ vs $\epsilon^{-1}$, $K=150$}
\end{subfigure}
\begin{subfigure}[b]{0.56\textwidth}
\centering
\includegraphics{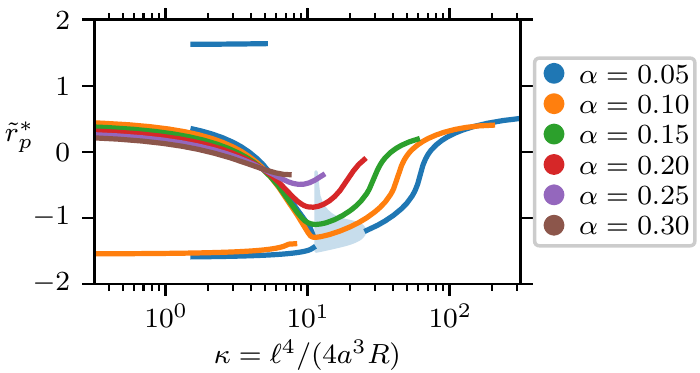}
\caption{$\tilde{r}_p^\ast$ vs $\kappa$, $K=150$}
\end{subfigure}
\caption{Horizontal location of stable equilibria $\tilde{r}_{p}^{\ast}$ versus (a,c,e) $\epsilon^{-1}$ and (b,d,e) $\kappa$, for the Dean numbers (a,b) $K=50$, (c,d) $K=100$, and (e,f) $K=150$. 
The duct cross-section has aspect ratio $2$ and $\tilde{r}_{p}^{\ast}$ is non-dimensionalised with respect to $\ell/2$. 
The light shaded area illustrates the region occupied by a stable orbit which occurs only when $\alpha=0.05$ for $K\gtrsim100$.
}\label{fig:focusing_collapse1}
\end{figure}

In Figure~\ref{fig:focusing_collapse1} we plot the horizontal focusing location of stable equilibria, denoted $\tilde{r}_{p}^{\ast}$, for the three Dean numbers $K=50,100,150$ in the case of a duct cross-section with aspect ratio $2$.
To facilitate comparison between different particle sizes, here $\tilde{r}_p^\ast$ has been non-dimensionalised with respect to half of the duct height.
The left plots show the change in $r_{p}^{\ast}$ versus $\epsilon^{-1}$ which is of practical interest when different size particles are suspended in flow through a duct having a specific bend radius.
The right plots show the change in $r_{p}^{\ast}$ versus $\kappa$ and illustrates how this parameter characterises the focusing behaviour.

The top row illustrates the case $K=50$ which is similar to the low flow rate results of \citet{HardingStokesBertozzi2019}, with just a slight shift towards the outside wall ($\tilde{r}=2$) of stable equilibria which focus near the centre ($\tilde{r}=0$).
We also include two additional particle sizes $\alpha=0.25,0.30$ here. 
The middle row shows the case of $K=100$ and, qualitatively, the results are very similar.
The main difference is that the increased flow rate shifts stable equlibria located near the centre (most noticeably those with $|\tilde{r}|\leq1/2$) further towards the outside wall ($\tilde{r}=2$).
Additionally, for a narrow range of $\epsilon^{-1}$, we observe that $\alpha=0.05$ particles no longer focus to stable equilibria but, instead, onto small stable orbits around unstable equilibria.
The bottom row shows the case of $K=150$.
The more pronounced shift in centrally focused equilibria is seen and there are subtle qualitative changes, particularly in relation to how sharply $\tilde{r}_p^\ast$ changes for larger $\kappa$ values.
Additionally, the stable orbits that occur for the particle size $\alpha=0.05$ now exist over a larger range of $\epsilon^{-1}$ and cover a greater portion of the cross-section.

The approximate collapse of focusing behaviour against $\kappa$ is preserved remarkably well as the Dean number $K$ is increased to a moderate size.
This result is significant in that is indicates that sized based separation of particles via inertial focusing should be reasonably robust to changes in flow rate.
Moreover we observe that particles focused closer to the centre of the duct shift towards the outer wall as the flow rate increases whereas those that are focused closer to the inside wall don't move quite so much.
This results in a slightly deeper trough in the plots of $r_{p}^{\ast}$ against $\kappa$ and further support the previous observation that separation is typically improved with increased flow rate.

As discussed in \citet{HardingStokesBertozzi2019}, the tail of data points where $\alpha=0.05$ and $\kappa<10$ is due to stable equilibria appearing near the centre of the inside and outside walls.
These occur for large $\epsilon^{-1}$ as the duct becomes straighter and secondary flow drag becomes negligible in comparison to the inertial lift force such that the equilibria of a straight rectangular duct are attained.
Similar stable equilibria near the centre of the inside wall exist for $\alpha=0.10$ and $\kappa<10$, although we expect these to ultimately disappear again in the limit $\kappa\to0$ (as a straight duct has no such equilibria).
The basin of attraction of these particular equilibria is very small and so it is reasonable to ignore them from a practical viewpoint (although were included here for completeness).

We briefly remark on the existence of stable orbits when $\alpha=0.05$ for sufficiently large $K$.
For these parameters there are no stable equilibria but instead a vertically symmetric pair of unstable (spiral) equilibria and around each of these there is a reasonably tight stable orbit.
These stable orbits occur over a relatively small parameter space for rectangular ducts in comparison to the case of square ducts whose dynamics are studied in~\citet{HaEtal22}. 
In this instance the required bend radius is $O(1000)$ times the duct height which is unlikely to occur in a practical setting.
Moreover, we expect that experimental noise would impact the stability of such orbits thereby making them difficult to observe experimentally.

\begin{figure}
\centering
\begin{subfigure}[b]{0.42\textwidth}
\centering
\includegraphics{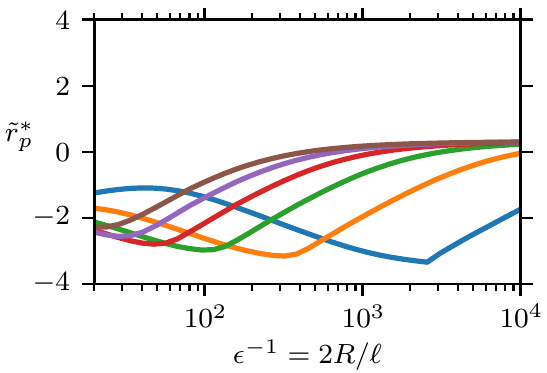}
\caption{$\tilde{r}_p^\ast$ vs $\epsilon^{-1}$, $K=50$}
\end{subfigure}
\begin{subfigure}[b]{0.56\textwidth}
\centering
\includegraphics{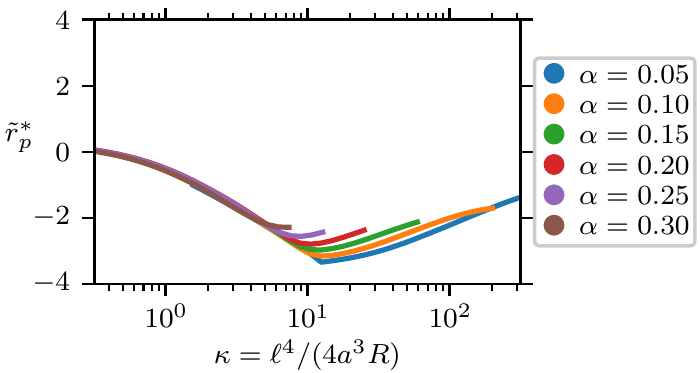}
\caption{$\tilde{r}_p^\ast$ vs $\kappa$, $K=50$}
\end{subfigure}
\\
\begin{subfigure}[b]{0.42\textwidth}
\centering
\includegraphics{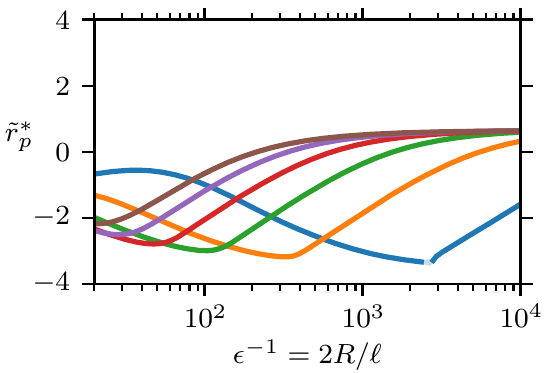}
\caption{$\tilde{r}_p^\ast$ vs $\epsilon^{-1}$, $K=100$}
\end{subfigure}
\begin{subfigure}[b]{0.56\textwidth}
\centering
\includegraphics{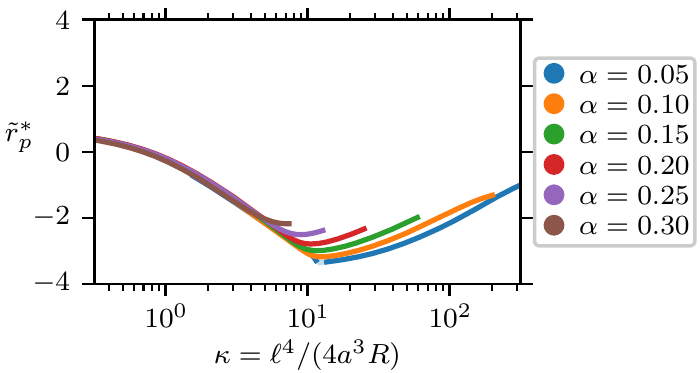}
\caption{$\tilde{r}_p^\ast$ vs $\kappa$, $K=100$}
\end{subfigure}
\\
\begin{subfigure}[b]{0.42\textwidth}
\centering
\includegraphics{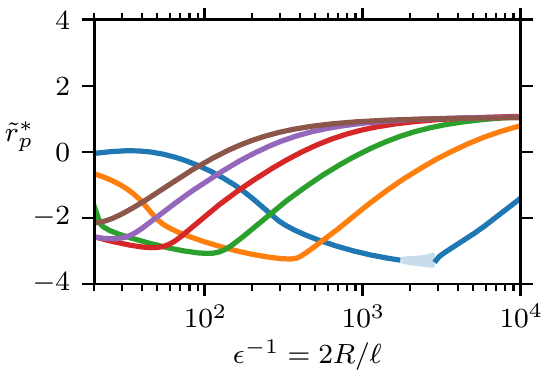}
\caption{$\tilde{r}_p^\ast$ vs $\epsilon^{-1}$, $K=150$}
\end{subfigure}
\begin{subfigure}[b]{0.56\textwidth}
\centering
\includegraphics{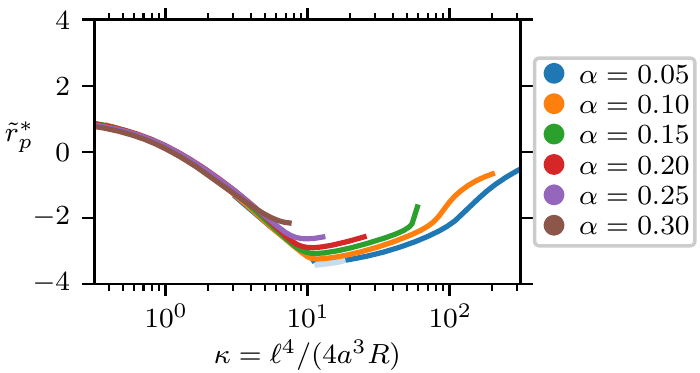}
\caption{$\tilde{r}_p^\ast$ vs $\kappa$, $K=150$}
\end{subfigure}
\caption{Horizontal location of stable equilibria $\tilde{r}_{p}^{\ast}$ versus (a,c,e) $\epsilon^{-1}$ and (b,d,e) $\kappa$, for the Dean numbers (a,b) $K=50$, (c,d) $K=100$, and (e,f) $K=150$. 
The duct cross-section has aspect ratio $4$ and $\tilde{r}_{p}^{\ast}$ is non-dimensionalised with respect to $\ell/2$. 
The light shaded area illustrates the region occupied by a stable orbit which occurs only when $\alpha=0.05$ for $K\gtrsim100$.
}\label{fig:focusing_collapse2}
\end{figure}

Figure~\ref{fig:focusing_collapse2} shows $r_{p}^{\ast}$ against $\epsilon^{-1}$ and $\kappa$ for a cross-section with aspect ratio $4$ and with Dean numbers $K=50,100,150$.
The main observation is that again centrally located equilibria shift closer to the outside wall with increasing $K$ which ultimately means separation can occur across a larger portion of the duct width.
The approximate collapse of the focusing curves against $\kappa$ for $\kappa<10$ is incredibly well preserved as $K$ increases.
The wider aspect ratio duct doesn't feature any stable equilibria appearing near the centre of the inside or outside wall, and stable orbits are very small where they do occur, thereby making the wider aspect ratio duct more attractive for size based separation.

\section{Conclusions}\label{sec:conclusions}

We have extended our model of inertial migration of neutrally buoyant spherical particles in curved microfluidic ducts to moderate Dean number by incorporating further terms of a suitable perturbation approximation of the background flow.
This approach is particularly powerful as the computed coefficient arrays can be cheaply re-assembled to model particle trajectories at any desired flow rate for which the model is applicable.
We have demonstrated a smooth shift in stable equilibria with respect to the Dean number $K$ and observed that increasing $K$ can enhance the separation of particles by size.
Moreover, we demonstrated that the approximate collapse of the horizontal location of equilibria when plotted against $\kappa$ is robust to increasing $K$.

We also investigated how long it takes particles to focus with respect to a time scale that accounts for the expected scaling with flow rate.
On this time scale, we found that results could be loosely separated into two distinct categories, faster focusing for $\kappa>10$, and slower focusing otherwise.
This can be explained by the existence of a slow manifold most particles must travel along when inertial lift is the dominant force.

Non-neutrally buoyant particles could easily be incorporated into this model by adding the appropriate perturbations as described in \citet{HardingStokes2020}.
We chose to focus on neutrally buoyant particles in this work to simplify the presentation and because our previous study concluded that small variations in buoyancy have negligible effect on particle dynamics.

A natural question going forwards is to understand how particle migration is perturbed by higher flow rates. 
This poses several challenges to the approach used herein.
First, the perturbation expansion of the background flow \eqref{eqn:ubar_pert} fails to converge for $K$ values larger than roughly $212$ \citep{HardingANZIAMJ2019}. 
One solution could be to expand around a non-zero $K$ value, or instead interpolate between solutions computed for several large $K$ values. 
Second, the assumption that $\Re_p$ remains suitable small becomes questionable and therefore additional terms in the perturbation expansion of the disturbance flow with respect to $\Re_p$ may need to be considered (although it is unclear if the radius of convergence for this expansion is large enough for this to be useful).
Third, it may no longer be appropriate to neglect the effects of particle acceleration, particularly for smaller particles which migrate rapidly around the secondary vortices.
All of these challenges will be considered going forwards as we continue to expand and improve our model of inertial migration in curved ducts.

\appendix

\section{Application of the reciprocal theorem to calculating hydrodynamic force and torque}\label{app:recip}

For completeness we begin by recalling the Lorentz reciprocal theorem for Stokes flow in its most general form and then revisit its application to the calculation of the force on a portion of the boundary of the fluid domain.
Following this we illustrate how it can also be applied to the calculation of the torque about a desired reference point over a portion of the boundary of the fluid domain.

\textbf{Theroem.}
Let $p,\mathbf{u}$ and $q,\mathbf{v}$ be the solution of 
\begin{subequations}\label{eqn:recip_setup}\begin{align}
\nabla\cdot\left(-p\mathbb{I}+\mu\big(\nabla\mathbf{u}+\nabla\mathbf{u}^{\intercal}\big)\right) &= \mathbf{f}\,, &\nabla\cdot\left(-q\mathbb{I}+\mu\big(\nabla\mathbf{v}+\nabla\mathbf{v}^{\intercal}\big)\right) &= \mathbf{g} &&\text{in $\Omega$}\,,\\
\nabla\cdot\mathbf{u} &= 0 \,,&\nabla\cdot\mathbf{v} &= 0 &&\text{in $\Omega$}\,, \\
\mathbf{u} &= \mathbf{a}\,, &\mathbf{v} &= \mathbf{b} &&\text{on $\partial\Omega$}\,,
\end{align}\end{subequations}
where $\mathbf{a},\mathbf{b},\mathbf{f},\mathbf{g}$ are suitably smooth vector fields over their respective domains, then
\begin{multline}\label{eqn:recip_general}
\int_{\partial\Omega}\mathbf{b}\cdot\left[\mathbf{n}\cdot\left(-p\mathbb{I}+\mu\big(\nabla\mathbf{u}+\nabla\mathbf{u}^{\intercal}\big)\right)\right]\,dS \\
= \int_{\partial\Omega}\mathbf{a}\cdot\left[\mathbf{n}\cdot\left(-q\mathbb{I}+\mu\big(\nabla\mathbf{v}+\nabla\mathbf{v}^{\intercal}\big)\right)\right]\,dS + \int_{\Omega}\mathbf{v}\cdot\mathbf{f}-\mathbf{u}\cdot\mathbf{g}\,dV \,,
\end{multline}
where $\mathbf{n}$ denotes the outward pointing normal vector with respect to $\Omega$.
\begin{proof}
Observe that \eqref{eqn:recip_setup} implies that
\begin{subequations}\label{eqn:recip_proof_1}\begin{align}
\mathbf{v}\cdot\left[\nabla\cdot\left(-p\mathbb{I}+\mu\big(\nabla\mathbf{u}+\nabla\mathbf{u}^{\intercal}\big)\right)-\mathbf{f}\right]&=0 \,, \\
\mathbf{u}\cdot\left[\nabla\cdot\left(-q\mathbb{I}+\mu\big(\nabla\mathbf{v}+\nabla\mathbf{v}^{\intercal}\big)\right)-\mathbf{g}\right]&=0 \,,
\end{align}\end{subequations}
over $\Omega$.
In general, given a differentiable tensor field $\mathbf{S}$, one has $\nabla\cdot(\mathbf{v}\cdot\mathbf{S})=\mathbf{v}\cdot(\nabla\cdot\mathbf{S})+\nabla\mathbf{v}:\mathbf{S}$ where $:$ is the tensor double dot product. 
Therefore \eqref{eqn:recip_proof_1} can be written as
\begin{subequations}\label{eqn:recip_proof_2}\begin{align}
\nabla\cdot\left[\mathbf{v}\cdot\left(-p\mathbb{I}+\mu\big(\nabla\mathbf{u}+\nabla\mathbf{u}^{\intercal}\big)\right)\right]-\nabla\mathbf{v}:\left(-p\mathbb{I}+\mu\big(\nabla\mathbf{u}+\nabla\mathbf{u}^{\intercal}\big)\right)
&=\mathbf{v}\cdot\mathbf{f} \,, \\
\nabla\cdot\left[\mathbf{u}\cdot\left(-q\mathbb{I}+\mu\big(\nabla\mathbf{v}+\nabla\mathbf{v}^{\intercal}\big)\right)\right]-\nabla\mathbf{u}:\left(-q\mathbb{I}+\mu\big(\nabla\mathbf{v}+\nabla\mathbf{v}^{\intercal}\big)\right)
&=\mathbf{u}\cdot\mathbf{g} \,.
\end{align}\end{subequations}
Subtracting the second equation from the first, then noting $\nabla\mathbf{v}: p\mathbb{I}=p(\nabla\cdot\mathbf{v})=0$ and $\nabla\mathbf{u}: q\mathbb{I}=q(\nabla\cdot\mathbf{u})=0$, and additionally $\nabla\mathbf{u}:\nabla\mathbf{v}=\nabla\mathbf{v}:\nabla\mathbf{u}$ and $\nabla\mathbf{u}:\nabla\mathbf{v}^{\intercal}=\nabla\mathbf{u}^{\intercal}:\nabla\mathbf{v}=\nabla\mathbf{v}:\nabla\mathbf{u}^{\intercal}$, then
\begin{equation}
\nabla\cdot\left[\mathbf{v}\cdot\left(-p\mathbb{I}+\mu\big(\nabla\mathbf{u}+\nabla\mathbf{u}^{\intercal}\big)\right)\right]
-\nabla\cdot\left[\mathbf{u}\cdot\left(-q\mathbb{I}+\mu\big(\nabla\mathbf{v}+\nabla\mathbf{v}^{\intercal}\big)\right)\right]
=\mathbf{v}\cdot\mathbf{f}-\mathbf{u}\cdot\mathbf{g} \,.
\end{equation}
It remains to integrate this equation over $\Omega$ and apply the divergence theorem to the left hand side.
The result is then straightforward to re-arrange to obtain \eqref{eqn:recip_general}.
\end{proof}

Now we consider the reciprocal theorem applied to the calculation of the hydrodynamic force from a flow $p,\mathbf{u}$ on a subset of $\partial\Omega$, e.g. denoting the surface of a particle. 
In particular, we consider calculating the component of the force
\begin{equation}
\int_{\Gamma}(-\mathbf{n})\cdot\left(-p\mathbb{I}+\mu\big(\nabla\mathbf{u}+\nabla\mathbf{u}^{\intercal}\big)\right) \,dS \,,
\end{equation}
on the surface $\Gamma\subset\partial\Omega$ projected onto a constant unit vector $\mathbf{e}_{\ast}$.

\textbf{Corollary.}
Let $p,\mathbf{u}$ and $q,\mathbf{v}$ satisfy \eqref{eqn:recip_setup} with $\mathbf{a}=\mathbf{0}$ over $\partial\Omega$, $\mathbf{g}=\mathbf{0}$ over $\Omega$, $\mathbf{b}=\mathbf{e}_{\ast}$ is a constant vector on $\Gamma\subset\partial\Omega$, and $\mathbf{b}=\mathbf{0}$ over the remainder of $\partial\Omega\backslash\Gamma$.
Then
\begin{equation}
\mathbf{e}_{\ast}\cdot\int_{\Gamma}(-\mathbf{n})\cdot\left(-p\mathbb{I}+\mu\big(\nabla\mathbf{u}+\nabla\mathbf{u}^{\intercal}\big)\right) \,dS
= -\int_{\Omega}\mathbf{v}\cdot\mathbf{f}\,dV \,,
\end{equation}
\begin{proof}
Immediately, due to $\mathbf{a}=\mathbf{0}$ and $\mathbf{g}=\mathbf{0}$, one obtains from \eqref{eqn:recip_general} the result
\begin{equation}
\int_{\partial\Omega}\mathbf{b}\cdot\left[\mathbf{n}\cdot\left(-p\mathbb{I}+\mu\big(\nabla\mathbf{u}+\nabla\mathbf{u}^{\intercal}\big)\right)\right]\,dS = \int_{\Omega}\mathbf{v}\cdot\mathbf{f}\,dV \,.
\end{equation}
Since $\mathbf{v}=\mathbf{b}=\mathbf{e}_{\ast}$ is constant over $\Gamma$ (and is zero elsewhere) it follows that
\begin{equation}
\mathbf{e}_{\ast}\cdot\int_{\Gamma}\mathbf{n}\cdot\left(-p\mathbb{I}+\mu\big(\nabla\mathbf{u}+\nabla\mathbf{u}^{\intercal}\big)\right) \,dS = \int_{\Omega}\mathbf{v}\cdot\mathbf{f}\,dV \,.
\end{equation}
One simply needs to multiply both sides by $-1$ to obtain the desired hydrodynamic force in the direction $\mathbf{e}_{\ast}$.
\end{proof}

Observe that in the corollary $\mathbf{v}$ could be written as $\boldsymbol{\mathcal{V}}(\mathbf{e}_{\ast})$ using the operator $\boldsymbol{\mathcal{V}}$ introduced in Section~\ref{sec:stokes_operators}, taking $\Gamma=\partial\hat{\mathcal{P}}'$.
Now, we wish to apply the reciprocal theorem again to the calculation of the torque 
\begin{equation}
\int_{\Gamma}(\mathbf{x}-\mathbf{x}_{c})\cross\left[(-\mathbf{n})\cdot\left(-p\mathbb{I}+\mu\big(\nabla\mathbf{u}+\nabla\mathbf{u}^{\intercal}\big)\right)\right] \,dS \,,
\end{equation}
about a (fixed) reference point $\mathbf{x}_{c}$, e.g. the centre of mass of a particle.

\textbf{Corollary.}
Let $p,\mathbf{u}$ and $q,\mathbf{v}$ satisfy \eqref{eqn:recip_setup} with $\mathbf{a}=\mathbf{0}$ over $\partial\Omega$, $\mathbf{g}=\mathbf{0}$ over $\Omega$, $\mathbf{b}=\mathbf{e}_{\ast}\cross(\mathbf{x}-\mathbf{x}_{c})$ on $\Gamma\subset\partial\Omega$ where $\mathbf{e}_{\ast}$ is constant, and $\mathbf{b}=\mathbf{0}$ over the remainder $\partial\Omega\backslash\Gamma$.
Then
\begin{equation}\label{eqn:recip_torque}
\mathbf{e}_{\ast}\cdot\int_{\Gamma}(\mathbf{x}-\mathbf{x}_{c})\cross\left[(-\mathbf{n})\cdot\left(-p\mathbb{I}+\mu\big(\nabla\mathbf{u}+\nabla\mathbf{u}^{\intercal}\big)\right)\right]\,dS
= -\int_{\Omega}\mathbf{v}\cdot\mathbf{f}\,dV \,,
\end{equation}
\begin{proof}
This time \eqref{eqn:recip_general} yields the result
\begin{equation}
\int_{\Gamma}\big(\mathbf{e}_{\ast}\cross(\mathbf{x}-\mathbf{x}_{c})\big)\cdot\left[\mathbf{n}\cdot\left(-p\mathbb{I}+\mu\big(\nabla\mathbf{u}+\nabla\mathbf{u}^{\intercal}\big)\right)\right]\,dS = \int_{\Omega}\mathbf{v}\cdot\mathbf{f}\,dV \,.
\end{equation}
Applying the triple product shift rule $(\mathbf{x}\cross\mathbf{y})\cdot\mathbf{z}=(\mathbf{y}\cross\mathbf{z})\cdot\mathbf{x}$ to the integrand on the left hand side produces
\begin{equation}
\int_{\Gamma}\left((\mathbf{x}-\mathbf{x}_{c})\cross\left[\mathbf{n}\cdot\left(-p\mathbb{I}+\mu\big(\nabla\mathbf{u}+\nabla\mathbf{u}^{\intercal}\big)\right)\right]\right)\cdot\mathbf{e}_{\ast}\,dS = \int_{\Omega}\mathbf{v}\cdot\mathbf{f}\,dV \,.
\end{equation}
The dot product with $\mathbf{e}_{\ast}$ can be safely pulled outside the integral (as $\mathbf{e}_{\ast}$ is constant).
Multiplying both sides by $-1$ then yields the desired component of torque in the $\mathbf{e}_{\ast}$ direction, i.e. \eqref{eqn:recip_torque}.
\end{proof}
Observe that using the operators introduced in Section~\ref{sec:stokes_operators} we can write $\mathbf{v}=\boldsymbol{\mathcal{V}}(\mathbf{e}_{\ast}\cross(\mathbf{x}-\mathbf{x}_{c}))$ within this corollary, taking $\Gamma=\partial\hat{\mathcal{P}}'$.

\section{Symmetries associated with Stokes flow around a spherical particle in a curved duct.}\label{app:symmetries}

Here we describe a variety of quantities which are zero due to symmetries associated with our problem.
Specifically, in the rotating frame the fluid domain $\mathcal{F}'$ is symmetric about $\theta'=0$ and many fields of interest relating to solutions of \eqref{eqn:vq_ast} are either odd or even with respect to $\theta'$.
Specifically, the following are odd with respect to $\theta'$
\begin{align}
&\boldsymbol{\mathcal{V}}(\vec{i}')\cdot\vec{j}' \,, &
& \boldsymbol{\mathcal{V}}(\vec{j}')\cdot\vec{i}' \,, &&\boldsymbol{\mathcal{V}}(\vec{j}')\cdot\vec{k} \,, \notag\\
& \boldsymbol{\mathcal{V}}(\vec{k})\cdot\vec{j}' \,, &
&\boldsymbol{\mathcal{V}}(\vec{i}'\cross(\hat{\vec{x}}'-\hat{\vec{x}}_{p}^{\prime}))\cdot\vec{i}'  \,, && \boldsymbol{\mathcal{V}}(\vec{i}'\cross(\hat{\vec{x}}'-\hat{\vec{x}}_{p}^{\prime}))\cdot\vec{k} \,, \notag\\
&\boldsymbol{\mathcal{V}}(\vec{j}'\cross(\hat{\vec{x}}'-\hat{\vec{x}}_{p}^{\prime}))\cdot\vec{j}'  \,, &
&\boldsymbol{\mathcal{V}}(\vec{k}\cross(\hat{\vec{x}}'-\hat{\vec{x}}_{p}^{\prime}))\cdot\vec{i}'  \,, && \boldsymbol{\mathcal{V}}(\vec{k}\cross(\hat{\vec{x}}'-\hat{\vec{x}}_{p}^{\prime}))\cdot\vec{k} \,, \notag\\
&\mathcal{Q}(\vec{j}') \,, && \mathcal{Q}(\vec{i}'\cross(\hat{\vec{x}}'-\hat{\vec{x}}_{p}^{\prime})) \,, && \mathcal{Q}(\vec{k}\cross(\hat{\vec{x}}'-\hat{\vec{x}}_{p}^{\prime})) \,, 
\end{align}
while the following are even with respect to $\theta'$
\begin{align}
&\boldsymbol{\mathcal{V}}(\vec{i}')\cdot\vec{i}' \,, && \boldsymbol{\mathcal{V}}(\vec{i}')\cdot\vec{k} \,, &
&\boldsymbol{\mathcal{V}}(\vec{j}')\cdot\vec{j}' \,, \notag\\
& \boldsymbol{\mathcal{V}}(\vec{k})\cdot\vec{i}' \,, && \boldsymbol{\mathcal{V}}(\vec{k})\cdot\vec{k} \,, &
&\boldsymbol{\mathcal{V}}(\vec{i}'\cross(\hat{\vec{x}}'-\hat{\vec{x}}_{p}^{\prime}))\cdot\vec{j}'  \,, \notag\\
& \boldsymbol{\mathcal{V}}(\vec{j}'\cross(\hat{\vec{x}}'-\hat{\vec{x}}_{p}^{\prime}))\cdot\vec{i}' \,, &&\boldsymbol{\mathcal{V}}(\vec{j}'\cross(\hat{\vec{x}}'-\hat{\vec{x}}_{p}^{\prime}))\cdot\vec{k}  \,, &
&\boldsymbol{\mathcal{V}}(\vec{k}\cross(\hat{\vec{x}}'-\hat{\vec{x}}_{p}^{\prime}))\cdot\vec{j}'  \,, \notag\\
&\mathcal{Q}(\vec{i}') \,, && \mathcal{Q}(\vec{k}) \,, && \mathcal{Q}(\vec{j}'\cross(\hat{\vec{x}}'-\hat{\vec{x}}_{p}^{\prime})) \,.
\end{align}
These properties can be used to show:
\begin{align}
\boldsymbol{\mathcal{N}}(\mathbf{i}')\cdot\mathbf{i}' &= 0 \,, & \boldsymbol{\mathcal{M}}(\mathbf{i}')\cdot\mathbf{j}' & = 0 \,, & \boldsymbol{\mathcal{N}}(\mathbf{i}')\cdot\mathbf{k} &= 0 \,, \notag\\
\boldsymbol{\mathcal{M}}(\mathbf{j}')\cdot\mathbf{i}' &= 0 \,, & \boldsymbol{\mathcal{N}}(\mathbf{j}')\cdot\mathbf{j}' &= 0 \,, &\boldsymbol{\mathcal{M}}(\mathbf{j}')\cdot\mathbf{k} &= 0 \,, \notag\\
\boldsymbol{\mathcal{N}}(\mathbf{k})\cdot\mathbf{i}' &= 0 \,, & \boldsymbol{\mathcal{M}}(\mathbf{k})\cdot\mathbf{j}' & = 0 \,, & \boldsymbol{\mathcal{N}}(\mathbf{k})\cdot\mathbf{k} &= 0 \,, \notag\\
\boldsymbol{\mathcal{M}}(\mathbf{i}'\cross(\hat{\vec{x}}'-\hat{\vec{x}}_{p}^{\prime}))\cdot\mathbf{i}' &= 0\,, & \boldsymbol{\mathcal{N}}(\mathbf{i}'\cross(\hat{\vec{x}}'-\hat{\vec{x}}_{p}^{\prime}))\cdot\mathbf{j}' &= 0 \,, & \boldsymbol{\mathcal{M}}(\mathbf{i}'\cross(\hat{\vec{x}}'-\hat{\vec{x}}_{p}^{\prime}))\cdot\mathbf{k} &= 0 \,, \notag\\
\boldsymbol{\mathcal{N}}(\mathbf{j}'\cross(\hat{\vec{x}}'-\hat{\vec{x}}_{p}^{\prime}))\cdot\mathbf{i}' &= 0 \,, & \boldsymbol{\mathcal{M}}(\mathbf{j}'\cross(\hat{\vec{x}}'-\hat{\vec{x}}_{p}^{\prime}))\cdot\mathbf{j}' & = 0 \,, & \boldsymbol{\mathcal{N}}(\mathbf{j}'\cross(\hat{\vec{x}}'-\hat{\vec{x}}_{p}^{\prime}))\cdot\mathbf{k} &= 0 \,, \notag\\
\boldsymbol{\mathcal{M}}(\mathbf{k}\cross(\hat{\vec{x}}'-\hat{\vec{x}}_{p}^{\prime}))\cdot\mathbf{i}' &= 0\,, & \boldsymbol{\mathcal{N}}(\mathbf{k}\cross(\hat{\vec{x}}'-\hat{\vec{x}}_{p}^{\prime}))\cdot\mathbf{j}' &= 0 \,, & \boldsymbol{\mathcal{M}}(\mathbf{k}\cross(\hat{\vec{x}}'-\hat{\vec{x}}_{p}^{\prime}))\cdot\mathbf{k} &= 0 \,.  \label{eqn:symm1} 
\end{align}
Moreover, given a sufficiently smooth function $f(r',z')$, i.e. which is independent of $\theta'$, then each of 
\begin{align}
&\boldsymbol{\mathcal{V}}(f(r',z')\vec{i}')\cdot\vec{j}' \,, & 
&\boldsymbol{\mathcal{V}}(f(r',z')\vec{j}')\cdot\vec{i}' \,, && \boldsymbol{\mathcal{V}}(f(r',z')\vec{j}')\cdot\vec{k} \,, \notag\\
&\boldsymbol{\mathcal{V}}(f(r',z')\vec{k})\cdot\vec{j}' \,, &
&\mathcal{Q}(f(r',z')\vec{j}') \,,
\end{align}
are odd with respect to $\theta'$, and each of 
\begin{align}
&\boldsymbol{\mathcal{V}}(f(r',z')\vec{i}')\cdot\vec{i}' \,, &&\boldsymbol{\mathcal{V}}(f(r',z')\vec{i}')\cdot\vec{k} \,, &
&\boldsymbol{\mathcal{V}}(f(r',z')\vec{j}')\cdot\vec{j}' \,, \notag\\
&\boldsymbol{\mathcal{V}}(f(r',z')\vec{k})\cdot\vec{i}' \,, &&\boldsymbol{\mathcal{V}}(f(r',z')\vec{k})\cdot\vec{k} \,, &
& \mathcal{Q}(f(r',z')\vec{i}') \,, && \mathcal{Q}(f(r',z')\vec{k}) \,,
\end{align}
are even with respect to $\theta'$. Consequently, one can show
\begin{align}
\boldsymbol{\mathcal{N}}(f(r',z')\mathbf{e}_{r})\cdot\mathbf{i}' &= 0 \,, & \boldsymbol{\mathcal{M}}(f(r',z')\mathbf{e}_{r})\cdot\mathbf{j}' & = 0 \,, & \boldsymbol{\mathcal{N}}(f(r',z')\mathbf{e}_{r})\cdot\mathbf{k} &= 0 \,, \notag\\
\boldsymbol{\mathcal{M}}(f(r',z')\mathbf{e}_{\theta})\cdot\mathbf{i}' &= 0\,, & \boldsymbol{\mathcal{N}}(f(r',z')\mathbf{e}_{\theta})\cdot\mathbf{j}' &= 0 \,, & \boldsymbol{\mathcal{M}}(f(r',z')\mathbf{e}_{\theta})\cdot\mathbf{k} &= 0 \,, \notag\\
\boldsymbol{\mathcal{N}}(f(r',z')\mathbf{k})\cdot\mathbf{i}' &= 0 \,, & \boldsymbol{\mathcal{M}}(f(r',z')\mathbf{k})\cdot\mathbf{j}' & = 0 \,, & \boldsymbol{\mathcal{N}}(f(r',z')\mathbf{k})\cdot\mathbf{k} &= 0 \,. \label{eqn:symm2} 
\end{align}
Of particular interest is when $f(r',z')$ described appropriate components of the background flow velocity field $\bar{\mathbf{u}}$, as described in Section~\ref{sec:bg_flow}.

\section{Further application of symmetry in the model}\label{sec:symmetry_reduction}

Similar to \citet{HardingStokesBertozzi2019}, we can further expand $\mathbf{I}_0$ to determine which terms contribute to and are most significant in \eqref{eqn:a1_vel_spin} and \eqref{eqn:s1_vel_spin}.
First, the decomposition leading to \eqref{eqn:leading_vel_spin} and \eqref{eqn:cs_vel_spin} suggests that $\mathbf{v}_{0}$ be split into two distinct parts, namely $\mathbf{v}_{0}=\mathbf{v}_{0,a}+\mathbf{v}_{0,s}$ where
\begin{subequations}\label{eqn:v0as_exp}\begin{align}
\mathbf{v}_{0,a} &= u_{p,0,\theta}\boldsymbol{\mathcal{V}}(\mathbf{j}')+\Omega_{p,0,r}\boldsymbol{\mathcal{V}}(\mathbf{i}'\cross(\hat{\vec{x}}'-\hat{\vec{x}}_{p}^{\prime}))+\Omega_{p,0,z}\boldsymbol{\mathcal{V}}(\mathbf{k}\cross(\hat{\vec{x}}'-\hat{\vec{x}}_{p}^{\prime})) \notag\\
&\quad-2\alpha^{-1}\left[\boldsymbol{\mathcal{V}}(\bar{\vec{u}}_{a,0})+K\boldsymbol{\mathcal{V}}(\bar{\vec{u}}_{a,0})+K^{2}\boldsymbol{\mathcal{V}}(\bar{\vec{u}}_{a,0})\right] \,, \\ 
\mathbf{v}_{0,s} &= u_{p,0,r}\boldsymbol{\mathcal{V}}(\mathbf{i}')+u_{p,0,z}\boldsymbol{\mathcal{V}}(\mathbf{k})+\Omega_{p,0,\theta}\boldsymbol{\mathcal{V}}(\mathbf{j}'\cross(\hat{\vec{x}}'-\hat{\vec{x}}_{p}^{\prime})) \notag\\&\quad-\kappa\Rey_{p}\left[\boldsymbol{\mathcal{V}}(\bar{\mathbf{u}}_{s,0})+K\boldsymbol{\mathcal{V}}(\bar{\mathbf{u}}_{s,1})+K^{2}\boldsymbol{\mathcal{V}}(\bar{\mathbf{u}}_{s,2})\right] \,.
\end{align}\end{subequations}
Subsequently, we can similarly decompose $\vec{I}_{0}$ into two distinct components, specifically $\vec{I}_{0}=\vec{I}_{0,a}+\vec{I}_{0,s}$ where
\begin{subequations}\label{eqn:I0as_exp}\begin{align}
\vec{I}_{0,a} &= \bs{\Theta}_{0}\cross\vec{v}_{0,a}+\vec{v}_{0,a}\cdot\nabla\left(2\alpha^{-1}(\bar{\vec{u}}_{a,0}+K\bar{\vec{u}}_{a,1}+K^2\bar{\vec{u}}_{a,2})\right) \\
&\quad +\left(\vec{v}_{0,a}+2\alpha^{-1}(\bar{\vec{u}}_{a,0}+K\bar{\vec{u}}_{a,1}+K^2\bar{\vec{u}}_{a,2})-\bs{\Theta}_{0}\cross\hat{\vec{x}}'\right)\cdot\nabla\vec{v}_{0,a} \notag\\
&\quad +\vec{v}_{0,s}\cdot\nabla\left(\kappa\Rey_p(\bar{\vec{u}}_{s,0}+K\bar{\vec{u}}_{s,1}+K^2\bar{\vec{u}}_{s,2})\right) \notag \\
&\quad +\left(\vec{v}_{0,s}+\kappa\Rey_p(\bar{\vec{u}}_{s,0}+K\bar{\vec{u}}_{s,1}+K^2\bar{\vec{u}}_{s,2})\right)\cdot\nabla\vec{v}_{0,s} \,, \notag\\
\vec{I}_{0,s} &= \bs{\Theta}_{0}\cross\vec{v}_{0,s}+\vec{v}_{0,s}\cdot\nabla\left(2\alpha^{-1}(\bar{\vec{u}}_{a,0}+K\bar{\vec{u}}_{a,1}+K^2\bar{\vec{u}}_{a,2})\right) \\
&\quad +\left(\vec{v}_{0,s}+\kappa\Rey_p(\bar{\vec{u}}_{s,0}+K\bar{\vec{u}}_{s,1}+K^2\bar{\vec{u}}_{s,2})\right)\cdot\nabla\vec{v}_{0,a} \notag\\
&\quad +\vec{v}_{0,a}\cdot\nabla\left(\kappa\Rey_p(\bar{\vec{u}}_{s,0}+K\bar{\vec{u}}_{s,1}+K^2\bar{\vec{u}}_{s,2})\right) \notag \\
&\quad +\left(\vec{v}_{0,a}+2\alpha^{-1}(\bar{\vec{u}}_{a,0}+K\bar{\vec{u}}_{a,1}+K^2\bar{\vec{u}}_{a,2})-\bs{\Theta}_{0}\cross\hat{\vec{x}}'\right)\cdot\nabla\vec{v}_{0,s}  \,. \notag
\end{align}\end{subequations}
Using the symmetries described in Appendix~\ref{app:symmetries} it can be shown that $\vec{I}_{0,a}\cdot\bs{\mathcal{V}}(\vec{j}')$ is odd with respect to $\theta$ such that
\begin{equation}
\int_{\hat{\mathcal{F}}'}\boldsymbol{\mathcal{V}}(\vec{j}')\cdot\mathbf{I}_{0} \,dV = \int_{\hat{\mathcal{F}}'}\boldsymbol{\mathcal{V}}(\vec{j}')\cdot\mathbf{I}_{0,s} \,dV \,.
\end{equation}
Analogous simplifications can be made for a number of other volume integrals over a dot product between $\mathbf{I}_0$ and a $\bs{\mathcal{V}}(\dots)$ term.

Upon applying these symmetry properties to \eqref{eqn:s1_vel_spin} and \eqref{eqn:s1_vel_spin} we arrive at
\begin{multline}\label{eqn:a1_vel_spin_exp}
A_a\begin{bmatrix} u_{p,1,\theta} \\ \Omega_{p,1,r} \\ \Omega_{p,1,z} \end{bmatrix}
=\begin{bmatrix}
\frac{8\pi}{3}\Theta_0 u_{p,0,r} \\[2pt]  
-\frac{8\pi}{15}\Theta_{0}\Omega_{p,0,\theta} \\[2pt] 
0 
\end{bmatrix}-\begin{bmatrix}
\vec{j}'\cdot\int_{\hat{\mathcal{P}}'}\bar{\vec{u}}_a\cdot\nabla\bar{\vec{u}}_s+\bar{\vec{u}}_s\cdot\nabla\bar{\vec{u}}_a\,dV \\[2pt]
\vec{i}'\cdot\int_{\hat{\mathcal{P}}'}(\hat{\vec{x}}'-\hat{\vec{x}}_{p}^{\prime})\cross(\bar{\vec{u}}_a\cdot\nabla\bar{\vec{u}}_s+\bar{\vec{u}}_s\cdot\nabla\bar{\vec{u}}_a)\,dV \\[2pt]
\vec{k}\cdot\int_{\hat{\mathcal{P}}'}(\hat{\vec{x}}'-\hat{\vec{x}}_{p}^{\prime})\cross(\bar{\vec{u}}_a\cdot\nabla\bar{\vec{u}}_s+\bar{\vec{u}}_s\cdot\nabla\bar{\vec{u}}_a)\,dV
\end{bmatrix} \\
+\begin{bmatrix}	
\int_{\hat{\mathcal{F}}'}\boldsymbol{\mathcal{V}}(\vec{j}')\cdot\mathbf{I}_{0,s} \,dV \\[2pt]
\int_{\hat{\mathcal{F}}'}\boldsymbol{\mathcal{V}}(\vec{i}'\cross(\hat{\vec{x}}'-\hat{\vec{x}}_{p}^{\prime}))\cdot\mathbf{I}_{0,s} \,dV \\[2pt]
\int_{\hat{\mathcal{F}}'}\boldsymbol{\mathcal{V}}(\vec{k}\cross(\hat{\vec{x}}'-\hat{\vec{x}}_{p}^{\prime}))\cdot\mathbf{I}_{0,s} \,dV
\end{bmatrix} \,,
\end{multline}
and similarly
\begin{multline}\label{eqn:s1_vel_spin_exp}
A_s\begin{bmatrix} u_{p,1,r} \\ u_{p,1,z} \\ \Omega_{p,1,\theta} \end{bmatrix}
=\begin{bmatrix} -\frac{4\pi}{3}\Theta_{0}^{2}(\hat{R}+\hat{r}_{p}) \\[2pt] 0 \\[2pt] \frac{8\pi}{15}\Theta_{0}\Omega_{p,0,r} \end{bmatrix}
-\begin{bmatrix}
\vec{i}'\cdot\int_{\hat{\mathcal{P}}'}\bar{\vec{u}}_a\cdot\nabla\bar{\vec{u}}_a+\bar{\vec{u}}_s\cdot\nabla\bar{\vec{u}}_s\,dV \\[2pt]
\vec{k}\cdot\int_{\hat{\mathcal{P}}'}\bar{\vec{u}}_a\cdot\nabla\bar{\vec{u}}_a+\bar{\vec{u}}_s\cdot\nabla\bar{\vec{u}}_s\,dV \\[2pt]
\vec{j}'\cdot\int_{\hat{\mathcal{P}}'}(\hat{\vec{x}}'-\hat{\vec{x}}_{p}^{\prime})\cross(\bar{\vec{u}}_a\cdot\nabla\bar{\vec{u}}_a+\bar{\vec{u}}_s\cdot\nabla\bar{\vec{u}}_s)\,dV
\end{bmatrix} \\
+\begin{bmatrix}
\int_{\hat{\mathcal{F}}'}\boldsymbol{\mathcal{V}}(\vec{i}')\cdot\mathbf{I}_{0,a} \,dV \\[2pt]
\int_{\hat{\mathcal{F}}'}\boldsymbol{\mathcal{V}}(\vec{k})\cdot\mathbf{I}_{0,a} \,dV \\[2pt]
\int_{\hat{\mathcal{F}}'}\boldsymbol{\mathcal{V}}(\vec{j}'\cross(\hat{\vec{x}}'-\hat{\vec{x}}_{p}^{\prime}))\cdot\mathbf{I}_{0,a} \,dV 
\end{bmatrix} \,.
\end{multline}
This decomposition illustrates a clear separation in how different parts of the leading order disturbance flow solution contribute to the hydrodynamic force generated by the first order disturbance flow solution.
In \eqref{eqn:a1_vel_spin_exp} and \eqref{eqn:s1_vel_spin_exp} $\bar{\vec{u}}_a$ should be interpreted as $\bar{\vec{u}}_{a,0}+K\bar{\vec{u}}_{a,1}+K^2\bar{\vec{u}}_{a,2}$ and similarly for $\bar{\vec{u}}_s$.
Both $\vec{v}_{0,a},\vec{v}_{0,s}$ from \eqref{eqn:v0as_exp} can be substituted into $\vec{I}_{0,a},\vec{I}_{0,s}$ in \eqref{eqn:I0as_exp}, and subsequently the volume integrals over $\hat{\mathcal{F}}'$ in \eqref{eqn:a1_vel_spin_exp} and \eqref{eqn:s1_vel_spin_exp} can be expressed as a polynomial function of $\vec{u}_{p,0},\bs{\Omega}_{p,0},\Theta_0,\kappa,\Re_p,K$ with variable coefficients depending on $r_p,z_p,\epsilon,\alpha$.
By sampling the coefficient fields over a suitable range of $r_p,z_p,\epsilon,\alpha$ and interpolating appropriately, we can efficiently estimate the values of $\vec{u}_{p,1},\bs{\Omega}_{p,1}$ given any value of $\vec{u}_{p,0},\bs{\Omega}_{p,0},\Theta_0,\kappa,\Re_p,K$.

\section*{Acknowledgements}
This research is supported under the Australian Research Council's Discovery Projects funding scheme (DP200100834) and a Future Fellowship (FT160100108) to YMS.
High performance computing resources provided by the R\=apoi HPC Cluster at Victoria University of Wellington and the Phoenix HPC service at the University of Adelaide were employed.

\section*{Declaration of Interests}
The authors report no conflicts of interest.


\end{document}